\makeatletter
\declare@file@substitution{revtex4-1.cls}{revtex4-2.cls}
\makeatother
\documentclass[twocolumn]{aastex631}

\usepackage{CJKutf8}

\newcommand{\cntext}[1]{\begin{CJK}{UTF8}{gbsn}#1\end{CJK}\kern-1ex}
\accepted{May 6, 2024}
\shorttitle{Microwave Oscillations}
\shortauthors{Sharma et al.}

\begin{document}

\correspondingauthor{Rohit Sharma}
\email{rohit.sharma@fhnw.ch}

\title{Study of Particle Acceleration using Fine Structures and Oscillations in Microwaves from Electron Cyclotron Maser}

\author[0000-0003-0485-7098]{Rohit Sharma}
\affiliation{University of Applied Sciences and Arts Northwestern Switzerland, 5210 Windisch, Switzerland}
\affiliation{Space, Planetary \& Astronomical Sciences \& Engineering (SPASE), Indian Institute of Technology, Kanpur, 208016, Uttar Pradesh, India}

\author[0000-0003-1438-9099]{Marina Battaglia}
\affiliation{University of Applied Sciences and Arts Northwestern Switzerland, 5210 Windisch, Switzerland}



\author[0000-0003-2872-2614]{Sijie Yu (\cntext{余思捷})} 
\affiliation{Center for Solar-Terrestrial Research, New Jersey Institute of Technology, 323 M L King Jr Blvd, Newark, NJ 07102-1982, USA}

\author[0000-0002-0660-3350]{Bin Chen (\cntext{陈彬})}
\affiliation{Center for Solar-Terrestrial Research, New Jersey Institute of Technology, 323 M L King Jr Blvd, Newark, NJ 07102-1982, USA}

\author[0000-0002-5431-545X]{Yingjie Luo (\cntext{骆英杰})}
\affiliation{School of Physics \& Astronomy, University of Glasgow, G12 8QQ, Glasgow, UK}

\author[0000-0002-2002-9180]{S\"am Krucker}
\affiliation{University of Applied Sciences and Arts Northwestern Switzerland, 5210 Windisch, Switzerland}
\affiliation{Space Sciences Laboratory, University of California, 7 Gauss Way, 94720 Berkeley, USA}

\begin{abstract}

The accelerated electrons during solar flares produce radio bursts and nonthermal X-ray signatures. The quasi-periodic pulsations (QPPs) and fine structures in spatial-spectral-temporal space in radio bursts depend on the emission mechanism and the local conditions, such as magnetic fields, electron density, and pitch angle distribution. Radio burst observations with high frequency-time resolution imaging provide excellent diagnostics. In converging magnetic field geometries, the radio bursts can be produced via the electron-cyclotron maser (ECM). Recently, using observations made by the Karl G. Jansky Very Large Array (VLA) at 1--2 GHz, \cite{Yu2023} reported a discovery of long-lasting auroral-like radio bursts persistent over a sunspot and interpreted them as ECM-generated emission. Here, we investigate the detailed second and sub-second temporal variability of this continuous ECM source. We study the association of 5-second period QPPs with a concurrent GOES C1.5-class flare, utilizing VLA's imaging spectroscopy capability with an extremely high temporal resolution (50 ms). We use the density and magnetic field extrapolation model to constrain the ECM emission to the second harmonic o-mode. Using the delay of QPPs from X-ray emission times, combined with X-ray spectroscopy and magnetic extrapolation, we constrain the energies and pitch angles of the ECM-emitting electrons to $\approx$4-8 keV and $>26^{\circ}$. Our analysis shows that the loss-cone diffusion continuously fuels the ECM via Coulomb collisions and magnetic turbulence between a 5 Mm--100 Mm length scale. We conclude that the QPP occurs via the Lotka-Volterra system, where the electron from solar flares saturates the continuously operating ECM and causes temporary oscillations.




\end{abstract}


\section{Introduction}

During a solar flare, a large amount of magnetic energy is dumped into the solar corona. A portion of this energy is responsible for the acceleration of particles, which travel away from the acceleration site, guided by surrounding magnetic fields. Some particles can get confined into magnetic traps, depending on the magnetic field configurations and the particle's energy and pitch angles \citep{Benz2017}. With time, these particles diffuse into the loss cone, eventually precipitating in the chromosphere \citep[e.g.][]{Melrose1982}. For a steady radio emission, the particle acceleration must balance diffusion into loss cone and loss \citep[e.g.][]{Melrose1976,Aschwanden1988I}. Most recently, \cite{Yu2023} discovered a radio burst event over a sunspot lasting for many hours. Interestingly, flare activities are located $\approx70''$ away from the radio source and showed no obvious temporal association with the radio bursts. They attributed the radio burst event to ECM emission. Here, the coronal volume above the sunspot possessed a converging magnetic topology necessary for hosting the ECM source and guide particle transport.

Unraveling magnetic topology is essential to understand particle transport. Evolving plasma and magnetic field loops rooted in the photosphere inhabit the solar corona. In sunspots, the magnetic field strengths can be as high as $\ge\!1000$ G. Broadly, the magnetic topology above sunspots consists of a fountain-like 3-D topology with a central high magnetic field umbra and a low magnetic field penumbra featuring outward directing magnetic fields on the periphery. In addition, many large-scale magnetic loops are commonly seen depending on the level of complexity of the sunspot/active region. The accelerated flare electrons follow these magnetic field features and travel large distances. For low thermal densities, the transport of the energetic electrons can fall into the collisionless regime. However, in denser coronal loops, collisions must be taken into consideration \citep[e.g.][]{Bai1982}. Particle propagation is further modified by the presence of turbulence in the magnetic loops, resulting in a more isotropic distribution of electron pitch angles. \citep{Kontar2011,Kontar2014,Musset2018}. The magnetic inhomogeneities and turbulence impact the radio wave propagation via wave scattering \citep[e.g.][]{Steinberg1971,Kontar2019,Sharma2020a}.


The accelerated electrons can emit X-rays via the thermal and nonthermal bremsstrahlung from their interactions with the ambient plasma in the chromosphere or corona. At radio wavelengths, emissions from the accelerated electrons can come from a variety of processes depending on the ambient plasma conditions, and the responsible emission mechanisms can be, e.g. thermal bremsstrahlung, gyro-synchrotron, and a variety of coherent emission mechanisms \citep{McLean1985}. Near strong magnetic field regions, especially near sunspots, the ECM is a favourable mechanism for producing radio emission from metric to microwave wavelengths under the right conditions \citep{Melrose1982,Robinson1991}. The ECM mechanism is excited by the accelerated electrons having positive gradient along the perpendicular velocity direction, i.e. $\partial f /\partial v_{\perp} > 0$ in the regions where $\omega_{pe}/\Omega_{ce} < 1$ \citep{Wu1979}. Here $v_{\perp}$ is the electron's velocity in the perpendicular direction to the magnetic field, $\omega_{pe}$ is the electron's plasma frequency and $\Omega_{ce}$ is the electron's gyroresonance frequency. Energetic electron distributions like loss-cone, horseshoe or ring etc. have been suggested to be responsible for inducing the instabilities to excite the emission. Such particle distributions can form in the coronal loops with the trapped electrons that originated in a solar flare. The ECM emission in solar context has been applied to a variety of radio bursts ranging from radio spikes \citep[e.g.][]{Sharma1984} to type-III bursts \citep[e.g.][]{Wu2005,Chen2017}. The growth of O- and X- modes depends on the coronal environment, more precisely on the ratio of plasma frequency ($\omega_e$) to the gyro-frequency ($\Omega_B$). In the coronal regions, where $\omega_{pe}/\Omega_{ce} < 1$ the growth rates of the $s=2$ X-mode and fundamental O- mode are higher than other modes \citep[e.g.][]{Sharma1984}. Such regions are formed near the high magnetic field regions, e.g. sunspots with converging magnetic field geometry. However, the timescale of the ECM instability growth is short compared to the observational capabilities \citep[e.g.][]{Melrose1982}. 

The ECM instability can produce prolonged radio sources either by repeated injections or the large diffusion timescales \citep{Aschwanden1988I}. Under quasi-linear relaxation, the trapped electrons or protons diffuse slowly for both O- and X-mode. The resultant timescales can vary from sub-second to many hours depending on the harmonic of the mode and details of the wave-particle interaction. The ECM sources, which are produced by the interplay of the driver source and quasi-linear relaxation, can produce pulsations due to wave-wave interactions \citep{Aschwanden1988II}. This phenomenon is one of the possible explanations for QPPs in radio bursts. We note that QPPs observed in other wavelengths, e.g. extreme-ultraviolet (EUV) and X-rays \citep{Sim2015} may be explained by multiple physical mechanisms \citep[e.g.][]{Nakariakov2009,Zimovets2021,McLaughlin2018,Kupriyanova2020}. For example, many QPPs seen in EUV can be associated with MHD processes like oscillations via sausage modes \citep{Kolotkov2015}. Sequential chromospheric evaporation can produce repetitive EUV loop top source \citep[e.g.][]{ Patsourakos2004,Sharma2016}, and the fast evaporation flows can produce turbulence in the looptop \citep{Ruan2019}. QPPs can also occur from the turbulence via kelvin-Helmholtz instability, leading to emission fluctuations \citep[e.g.][]{Ruan2019}. 
X-ray and radio imaging of the QPPs provide greater depth in the involved physical and emission processes and even magnetography of the solar flaring loops \citep{Gary2013}. \cite{Luo2022} observed a QPP source at the looptop and observed concurrent sources in different loops but with different emission mechanisms. Previously, such concurrent sources with different origins in solar flares were observed by \citep[e.g.][]{Sharma2020}, however, not in QPPs. \cite{Kou2022} observed two QPP microwave sources with differing brightness temperatures ($T_{B}$) and periods of about 10-20 s and 30-60 s, respectively, and attributed them to arising from the modulation of magnetic islands. The timescales of the radio emission possess finer structures down to millisecond levels \citep[e.g.][]{Wang1999,2006AIPC}.
The other mechanisms include periodic magnetic reconnections, loop coalescence, and thermal dynamical cycles \citep{McLaughlin2018,Clarke2021ApJ,Nakariakov2006}. QPPs and solar flares can show interesting correlations, for e.g. between the QPP period and flare duration\citep{Hayes2020ApJ}. 
However, a coherent relationship between the ECM radio emission, QPPs and the energetic electrons is not well-established due to the lack of radio imaging with sub-second resolution, especially in the context of a turbulent coronal medium.

\begin{figure*}
\begin{center}
\begin{tabular}{c}
\resizebox{140mm}{!}{
\includegraphics[trim={0.0cm 0cm 0.0cm 0.0cm},clip,scale=0.5]{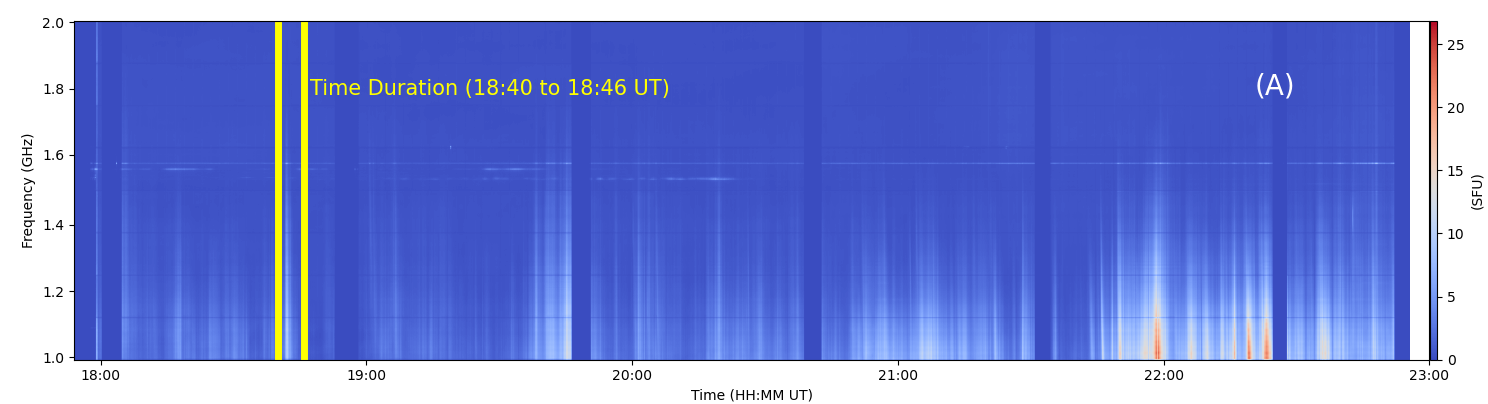}} \\
\resizebox{140mm}{!}{
\includegraphics[trim={0.0cm 0cm 0.0cm 0.0cm},clip,scale=0.5]{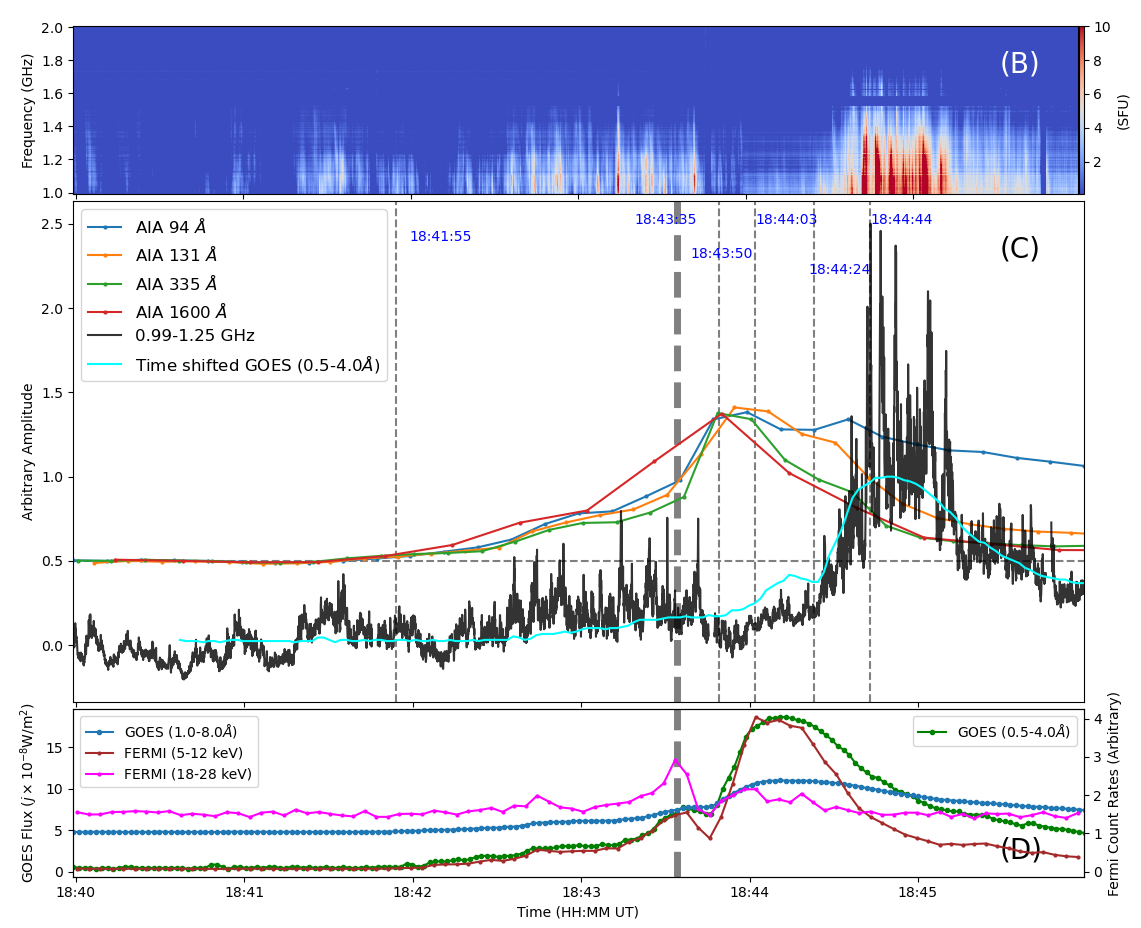}} 
\end{tabular} 
\end{center}
\caption{ Panel A: VLA total power dynamic spectrum of radio emission from 1-2 GHz for 5 hour time period and RCP polarisation on 16 April 2016. The region in between the yellow lines shows the radio burst of interest. Panel B: Zoom-in total power spectrum from panel A, i.e. in between the yellow vertical lines for the time of interest of the radio burst. Some bad channels have been flagged between 1.5 GHz and 1.7 GHz. 
Panel C: Time series of various AIA bands obtained by taking a spatial average over the flaring region marked in the magenta spot in Fig. \ref{fig:aia} (B) at (-810", 205"). The blue, orange, green, and red show the wavelength 94 \AA, 131 \AA, 335 \AA, 1600 \AA\ respectively. The black curve shows radio emission timeseries by averaging frequency bands from 1.0 to 1.25 GHz. Note that the amplitudes are arbitrary and scaled to highlight the temporal association with EUV and X-ray emission. The thin gray vertical dashed lines mark the start of EUV emission (18:41:55), the peak of EUV (18:43:50), the peak of thermal X-rays (18:44:03), radio burst start time (18:44:24) and radio burst peak (18:44:44), while bold dashed gray line shows the nonthermal X-ray peak at 18:43:35. The cyan curve is the GOES 0.5-4.0\AA\ forwardly time shifted by 38 sec. Note that all curves have arbitrary amplitudes. Panel D: GOES light curves for low- and high-energy bands are shown in blue and green, respectively. The GOES y-axis ($j\times 10^{-8}$ W m$^{-2}$) values have multiplicative factors $j=100$ and $j=1$ for 1.0-8.0 \AA\ and 0.5-4.0 \AA\ respectively. The brown and magenta curve shows the time series from Fermi/GBM's X-ray emission for 5-12 and 18-28 keV energy bands respectively. 
}
\label{Fig:ds}
\end{figure*}

Here, we use radio imaging spectroscopy with a fine temporal resolution from VLA, X-ray spectroscopy observations from Fermi's Gamma-ray Burst Monitor (GBM) \citep{Atwood2009}, EUV images from the Atmospheric Imaging Assembly \citep[AIA;][]{Lemen2012} on board the Solar Dynamics Observatory \citep[SDO;][]{Pesnell2012}, and magnetic field modeling to explore the details of the radio ECM source and the associated QPP phenomenon. We explore second and sub-second variations from the long-lasting radio emission over a sunspot reported by \cite{Yu2023}. Combining VLA's high-time resolution capability within the holistic picture from multiwavelength analysis and realistic coronal models allows us to study the temporal variations of ECM in exceptional detail. The observational overview is given in Section \ref{sec:observation}. The X-ray and radio data analysis are presented in Section \ref{sec:data_analysis}. 
The spatial-spectral-temporal characteristics and analysis of the radio emission and emission mechanism are discussed in Section \ref{sec:sst}. An electron propagation model for the ECM emission is presented in Section \ref{sec:model_ecm}. An analysis of QPPs in ECM is discussed in Section \ref{sec:pulsations}, followed by an analysis of fine structures in Section \ref{sec:fine}. Later, we discuss the findings in Section \ref{sec:discussion}, followed by conclusions in Section \ref{subsec:conclusion}.



\section{Observations}
\label{sec:observation}

The event of interest, the C1.5-class solar flare, occurred on 2016 April 9, from 18:40 UT to 18:46 UT and peaked at 18:44 UT. We study the flare using the radio data from the VLA, GBM X-rays, and AIA's EUV wavelengths. The day is marked by high solar activity with nine C-class flares, including the one under study, all in the active region 12529 positioned at N10E52  that hosts a $\beta$-class sunspot. Figure \ref{Fig:ds}(A) shows the VLA total power dynamic spectrum for the 5-hour period from 17:55 to 22:55 UT. It is obtained by median-averaging over all baselines from VLA's L-band receiver for right-hand circular polarization (RCP). Note that the entire VLA observation period is marked by low-frequency dominated radio bursts. \cite{Yu2023} investigate the long-term variation over many hours and attributes to auroral-like radio emission over the sunspot. Here, we focus on finer variations of the emission. Figure \ref{Fig:ds}(B) shows a zoomed-in dynamic spectrum (DS) of the SOL20160409T18:44 event. We note that the radio emission occurs from 18:44 to 18:46 UT, dominantly between the 1.0 and 1.5 GHz frequency range. There are also fainter spiky emissions between 18:42 and 18:44 UT preceding the stronger radio emission. The emission fluctuated but was steady before 18:44 UT, while brighter and more pronounced later. Therefore, we classify the period from 18:44 UT to 18:46 UT as radio bursts. The radio bursts in RCP are highly polarised ($\sim 70\%$). The radio bursts are bright with the peak flux densities $\approx$18 Solar Flux Units (SFU; 1 SFU =$10^{-22}$ J m$^{-2}$ Hz$^{-1}$ s$^{-1}$) at 1 GHz. They are brighter at low frequencies and fainter at higher frequencies. As the VLA observation did not cover frequencies below 1 GHz, although the bursts may likely extend to lower frequencies, they will not be analyzed in this study. The radio bursts are impulsive in time and show a QPP on the second timescale. Some fainter bursts can also be seen in prior times before 18:44 UT in the low frequencies in a lower frequency range (1.0 to 1.3 GHz) down to maximum flux density $\sim 10$ SFUs. 

\subsection{Event Outline}
\label{subsec:overview}

\begin{figure*}
    \centering
    \begin{tabular}{cccc}
    \resizebox{70mm}{!}{
\includegraphics[trim={0.0cm 0cm 0.0cm 0.0cm},clip,scale=0.6]{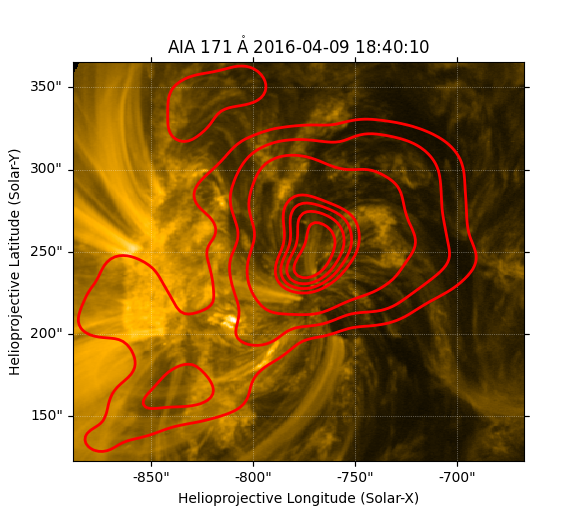}} &
        \resizebox{65mm}{!}{
\includegraphics[trim={0.0cm 0cm 0.0cm 0.0cm},clip,scale=0.4]{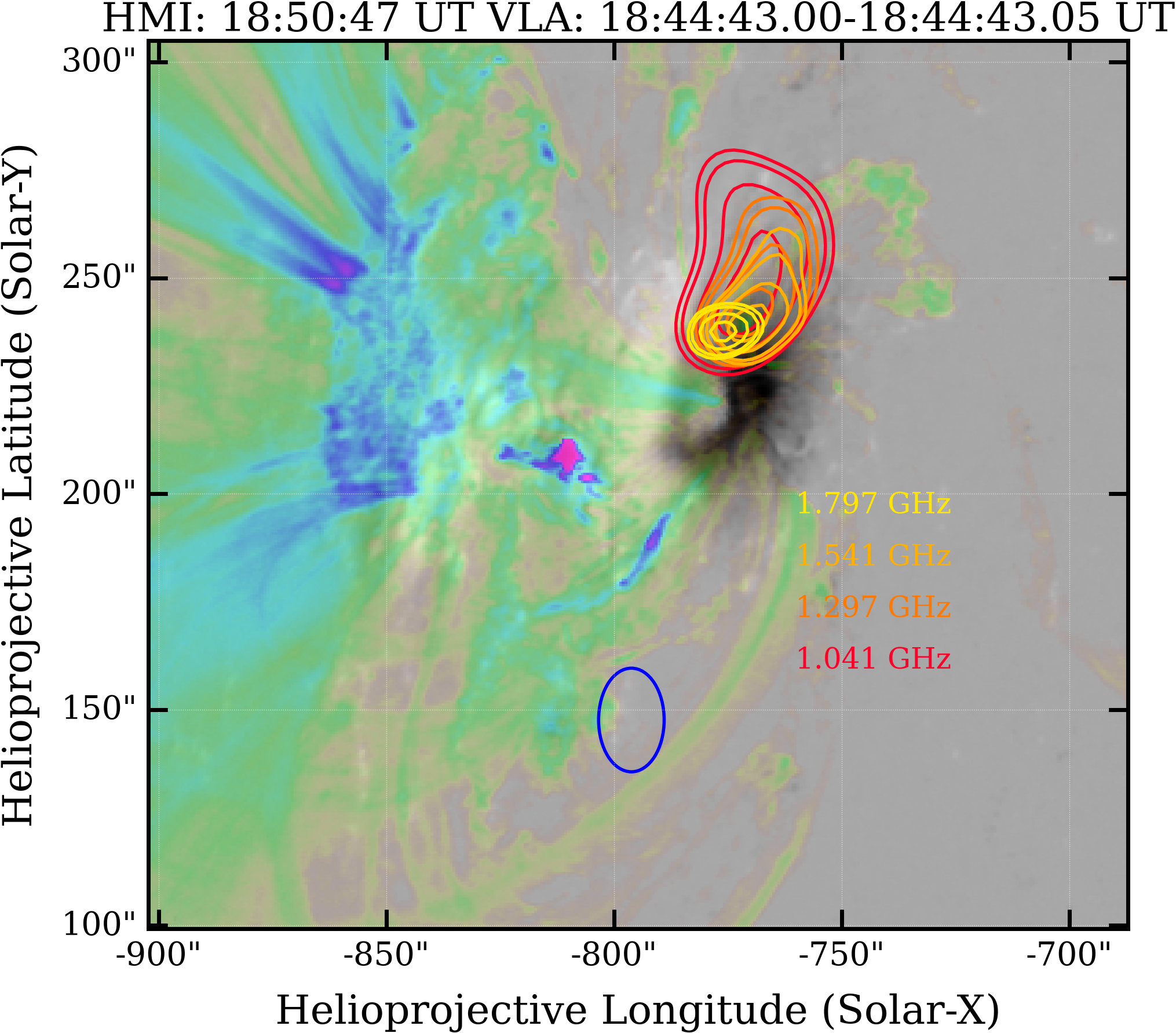}}  \\
(A) AIA 171$\AA$ & (B) Composite map \\
\resizebox{70mm}{!}{
\includegraphics[trim={0.0cm 0cm 0.0cm 0.0cm},clip,scale=0.6]{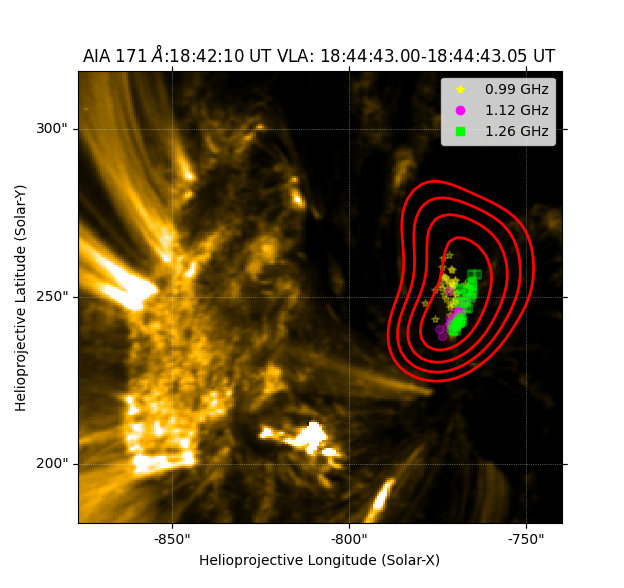}} &
\resizebox{79mm}{!}{
\includegraphics[trim={0.0cm 0cm 0.0cm 0.0cm},clip,scale=0.6]{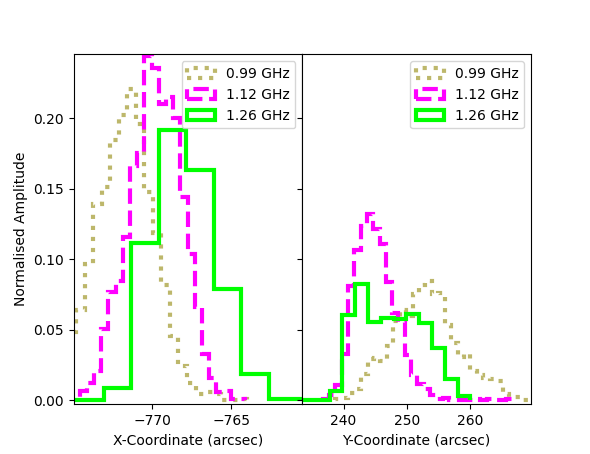}}\\
 (C) Radio centroids & (D) Centroid histograms
    \end{tabular}
    \caption{Panel A: Background image is AIA 171$\AA$ along with 1 GHz VLA radio source contours during continuum time at 18:40:10 UT. The radio contours levels are at 5\%, 10\%, 20\%, 40\%, 60\%, 70\%, 80\% and 90\% w.r.t image's maximum $T_B$ ($\approx$ 40 MK). Panel B: HMI magnetogram,  radio burst contours along with AIA 171 $\AA$ in transparent blue-magenta colormap. Radio burst contours are shown for four frequencies at 60\%, 70\%, 80\%, and 90\% w.r.t respective maximum $T_B$. We note that the burst source is extended at lower frequencies while compact at higher frequencies. The blue ellipse in the middle shows the size of the synthesized radio beam at 1 GHz. Panel C: AIA 171 \AA\ image with 40 random radio centroid locations between 18:44-18:46 UT, and 1 GHz radio burst contours (red) at 18:44:43 UT. The random locations were chosen to avoid over-crowding of the centroids. The centroid locations are shown for three frequencies, i.e. 0.99 GHz, 1.12 GHz, and 1.26 GHz. We note that the centroid locations for all sources are compact. The radio burst contour levels are at 60\%, 70\%, 80\% and 90\% w.r.t image's maximum, where maximum $T_B\approx$ 110 MK at 18:44:43 UT. Panel D: Normalised histograms of the X- and Y-coordinates of all the radio centroids. Note that the histogram extremities are within one primary beam.
    }
    \label{fig:aia}
\end{figure*}

The multi-wavelength behaviour of the solar flare is captured in Fig. \ref{Fig:ds} (C \& D). Panel D shows the detected X-ray emission from the Fermi/GBM X-ray for 5--12 keV (low-energy) and 18-28 keV (high-energy) and GOES wavabands. In panel C, the VLA light curve shows the spiky nature of the radio emission and is obtained by averaging the DS over frequency. The light curve leading to the radio burst begins to rise from 18:44:24 UT, i.e. start time of the radio burst. In panel C, the AIA light curves are obtained by averaging the flaring region shown as the magenta spot in Fig. \ref{fig:aia}(B) for each image in time. The rising phase of the AIA light curve starts at $\sim$ 18:41:55 UT, reaches the peak at $\sim$ 18:43:50 UT and decays thereafter, while the radio burst peaks at 18:44:44 UT, i.e. 54 sec later. We also note that the peak of the high energy X-rays does not coincide with either EUV or radio and precedes the AIA light curve peak at 18:43:50 UT by $\sim$28 sec. However, the low-energy X-ray peaks at 18:44:03 UT shortly after the AIA peak by 13 sec. Most notably, the start of the radio bursts occurs at 18:44:24, $\sim$ 49 sec and 21 sec later than the high and low energy Fermi/GBM X-ray peaks, respectively.
Overall, the time lags between the burst's start time and AIA EUV, Fermi/GBM low- and high-energy X-ray peaks are $\sim$ 34, 21, and 49 sec respectively. 

The GOES X-ray light curve at 1--8 \AA\ and 0.5--4.0 \AA\ in panels (D) shows the X-ray continuum of the event. We note that GOES 0.5-4.0 \AA\ closely resembles low-energy X-ray FERMI/GBM lightcurve, while the radio bursts (panel C) appear to be delayed with respect to GOES 0.5-4.0 \AA\ curve. By visual inspection, we delayed the time axis by 38 sec (call it $t_{delay}$) and plotted the time-shifted GOES curve in panel C. The GOES and radio burst profile match temporally, including a small peak between 18:44:00 to 18:44:24 UT. This small peak has a delay of 38 sec with respect to (w.r.t.) the high-energy X-ray peak.

\section{Data Analysis}
\label{sec:data_analysis}
Here, we discuss the results of radio imaging spectroscopy for the event using VLA L-band data and X-ray spectral analysis of the associated solar flare. Further, we also perform magnetic extrapolation modelling for the active region and the sunspot. Using them along with density models, we constrain the conditions for ECM. Finally, we analyze the observed radio source in the framework of the magnetic and density models and discuss it in detail.

\subsection{Radio Analysis and Imaging}
\label{sec:radio}
For the VLA observing program (VLA/16A-377), the VLA observation had L-band frequency coverage (1-2 GHz) with a spectral resolution of 1 MHz and temporal resolution of 50 ms with dual RCP and LCP polarization. The VLA observations were taken in sub-array mode in configuration C. In the sub-array mode, 14 antennas were used to observe in the L-band, while the rest 13 were in the S-band (2-4 GHz). Since the radio bursts occur mostly between 1.0 to 1.5 GHz, we will ignore the S-band in this paper. The L-band sub-array had a maximum baseline of 3.06 km and produced a synthesised beam of $\approx$15" at 1 GHz. The beam size further scales inversely with the frequency ($1/\nu$).

We produce radio maps for a frequency average of 4 MHz and a time resolution of 50 ms. Therefore, a total of 76,800 radio maps were produced from 18:44:00 UT to 18:46:00 UT. In addition, we also produced radio maps for one channel at 1 GHz from 18:40:00 to 18:44:00 UT to study the emission before the radio burst. We perform standard radio data reduction and synthesis imaging steps, including flagging, calibration, and
deconvolution using CASA \citep{McMullin2007}. For this observation, we use 3C48 as the flux and bandpass calibrator and the phase calibrator. An attenuation of 20 dB was applied on the solar scans in the signal path to reduce the
antenna gain, and corrections of their phase and amplitude have been applied following the method in \citet{Chen2013PhDT}. The obtained radio maps were made using \texttt{suncasa}\footnote{https://github.com/suncasa} for comparison with EUV and X-ray maps. The values in the radio maps were brightness temperature ($T_B$) obtained using the same prescription mentioned in \cite{Sharma2020}. Fig. \ref{fig:aia}(A-C) shows the radio source contours during the continuum phase (A) and burst (B \& C). Panel C also shows the location of the centroids for all radio sources for the 2 minutes of analysis (18:44 - 18:46 UT). This is obtained by fitting 90\% contour w.r.t the brightness peak of the respective image with an ellipse. The centroid location of the fitted ellipses is shown. We note that all the radio sources originate from a compact region above the sunspot. A more detailed discussion can be found later in Section \ref{subsec:ecm}.

The total power (Fig. \ref{Fig:ds} (B)) captures the integrated flux density of the radio source and includes a spatial average quantity over the source extent. Since our source is resolved by the instrument PSF, especially at low frequencies, the radio source model would consist of multiple superposed sources. To best capture the temporal variability of the radio source, the maximum brightness of the radio source is a better estimate than any spatially averaged quantity, like integrated flux density. We note that the derived brightness temperatures, $T_B$, are contingent upon the observed source size, which is at least as large as the synthesized beam ($12'' \nu_\mathrm{GHz}^{-1}$). Crucially, the observed source size might be several orders of magnitude larger than the actual intrinsic size of the source. Consequently, the intrinsic brightness temperature could potentially be orders of magnitude higher than the derived $T_B$ values. This significant difference is discussed in \citep{Yu2023}.

\subsection{Fermi/GBM X-ray Spectrum}
\label{subsec:fermi}
\begin{figure}
    \centering
\begin{tabular}{c}
\resizebox{80mm}{!}{
\includegraphics[trim={0.0cm 0cm 0.0cm 0.0cm},clip,scale=0.6]{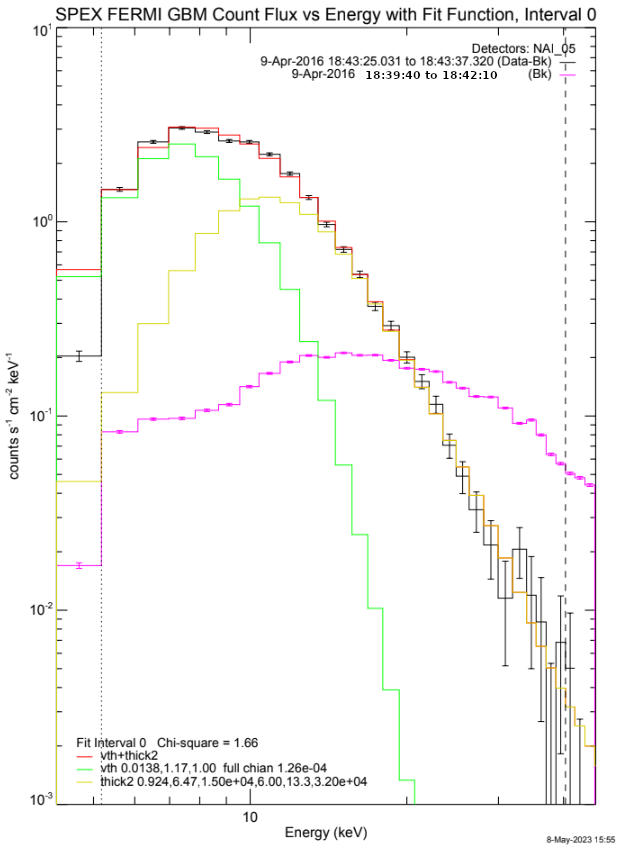}} 
\end{tabular}
    \caption{Fermi/GBM spectrum along with the fitted thermal and nonthermal components between 18:43:25 UT and 18:43:37 UT. The black datapoints show the Fermi/GBM spectrum, while the red curve shows the sum of the thermal (green line) and nonthermal (yellow line) components. The magenta histogram shows the background spectrum for 18:39:09 to 18:40:30 UT.}
    \label{fig:fermi}
\end{figure}

We fit the Fermi/GBM X-ray spectrum between 18:43:25 UT and 18:43:37 UT, around the high-energy X-ray peak using the standard OSPEX package \citep{Smith2002} shown in Figs. \ref{Fig:ds} \& \ref{fig:fermi}.
The spectral fit assumes a thermal model for the low energies and a nonthermal power-law component for higher energies under the thick-target bremsstrahlung regime (denoted as ``vth+thick2"). Fig. \ref{fig:fermi} shows the observed and fitted spectrum along with the background. The background emission computed from 18:39:09 UT to 18:40:30 UT was subtracted. The fit results suggest that a weak nonthermal emission was present during the high energy X-ray peak ($\sim$18:43:25--18:43:37 UT).

The fitted parameters are listed in Tab. \ref{tab:fermi}. We note that a weak nonthermal component is present in the spectrum with sufficient signal-to-noise until $\approx$ 30 keV with an electron spectral index $\delta$ of 6.4 and a low-energy cutoff ($E_{cut}$) of 13.25 keV and total electron flux of $9.2\pm2.6\times10^{34} s^{-1}$
($F_e (s^{-1})$; i.e., the total number of nonthermal electrons above the cutoff energy $E_{cut}$ per unit time). The electron nonthermal density $n_{nth}$ is given by  $n_{nth} = \frac{F_e}{v_e A}$,
where $v_e$ and $A$ are the electron beam velocity and flare footpoint area respectively. We can use $E_{cut}$ to estimate $v_e = \sqrt{\frac{2 E_{cut}}{m_e}} \approx 6.8 \times 10^{9}$ cm/s. For the flare footpoint area, we use the 171 $\AA$ EUV flare bright spot shown in magenta in Fig. \ref{fig:aia} (B) around (-810", 205"), i.e., $A \approx 10"\times10" = 5.27 \times 10^{17}$ cm$^{2}$. Using them, we get $n_{nth} \approx (2.57 \pm 0.85)\times 10^{7}$ cm$^{-3}$. 
Further, the nonthermal power in the nonthermal electrons is given by
\begin{equation}
P_e = (\frac{\delta -1}{\delta-2}) F_e E_{cut}.
\end{equation}
Putting in the above-calculated values, we obtain the total nonthermal power as $P_e \approx (2.4 \pm 0.7) \times 10^{27}$ ergs s$^{-1}$. While the thermal density is given by, $n_{th,l} = \sqrt{EM/V}$ for a source volume $V$. We approximate the source volume $V\sim A^{1.5}$, i.e $V\approx3.8\times10^{26}$ cm$^{3}$ and $n_{th,x}\approx 1.2 \times 10^{10}$ cm$^{-3}$. The subscript $x$ in $n_{th}$ indicates thermal density derived near the X-rays deposition site. Note that this footpoint area will be larger than the actual X-ray source due to its coronal origin, i.e. true thermal density will be larger, and the calculated value is a lower limit. 

\begin{table}[]
    \centering
    \begin{tabular}{|c|c|}
    \hline 
     Fermi/GBM Parameters    &  Values\\
     \hline \hline
        Electron flux ($F_e$) & $(9.2\pm2.6) \pm  \times 10^{34}$ s$^{-1}$ \\
        Electron spectral index ($\delta$) & 6.4$\pm$1.4 \\
        Emission Measure & $(1.4 \pm 0.1) \times 10^{47}$ cm$^{-5}$ \\
        Thermal density ($n_{th,x}$) & (6.1 $\pm$ 0.1)$\times 10^9$ cm$^{-3}$ \\
        Nonthermal density ($n_e$) & $(2.57 \pm 0.85)\times 10^{7}$ cm$^{-3}$ \\
        Low Energy Cut-off ($E_{cut}$) & 13.25 $\pm$ 0.62 keV \\
        Nonthermal Power (P$_e$) & $(2.4 \pm 0.7) \times 10^{27}$ ergs s$^{-1}$ \\
    \hline
    \end{tabular}
    \caption{Thermal and nonthermal X-ray spectral fit parameters done between 18:43:25 to 18:43:37 UT. The spectrum is shown in Fig. \ref{fig:fermi}.}
    \label{tab:fermi}
\end{table}


\section{An Overview of Emission Features and Mechanism}
\label{sec:sst}

In this section, we discuss the evolution of radio emission w.r.t EUV flare evolution and constraining radio emission mechanism and heights.
\subsection{Spatial Features}
\label{subsec:sst_overview}

\begin{figure*}
    \centering
    \resizebox{120mm}{!}{
\includegraphics[trim={0.0cm 0cm 0.0cm 0.0cm},clip,scale=0.6]{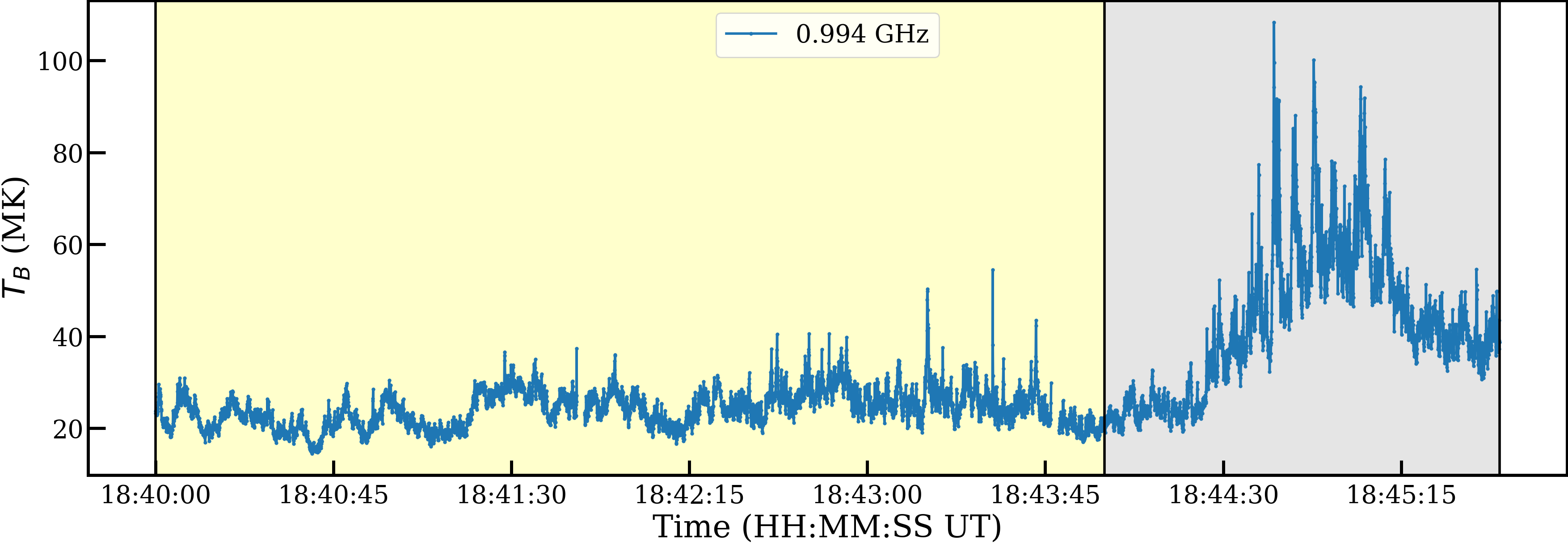}}\\
(A) 1 GHz $T_B$ timeseries  \\
    \begin{tabular}{ccc}
   \resizebox{55mm}{!}{
\includegraphics[trim={0.0cm 0cm 0.0cm 0.0cm},clip,scale=0.3]{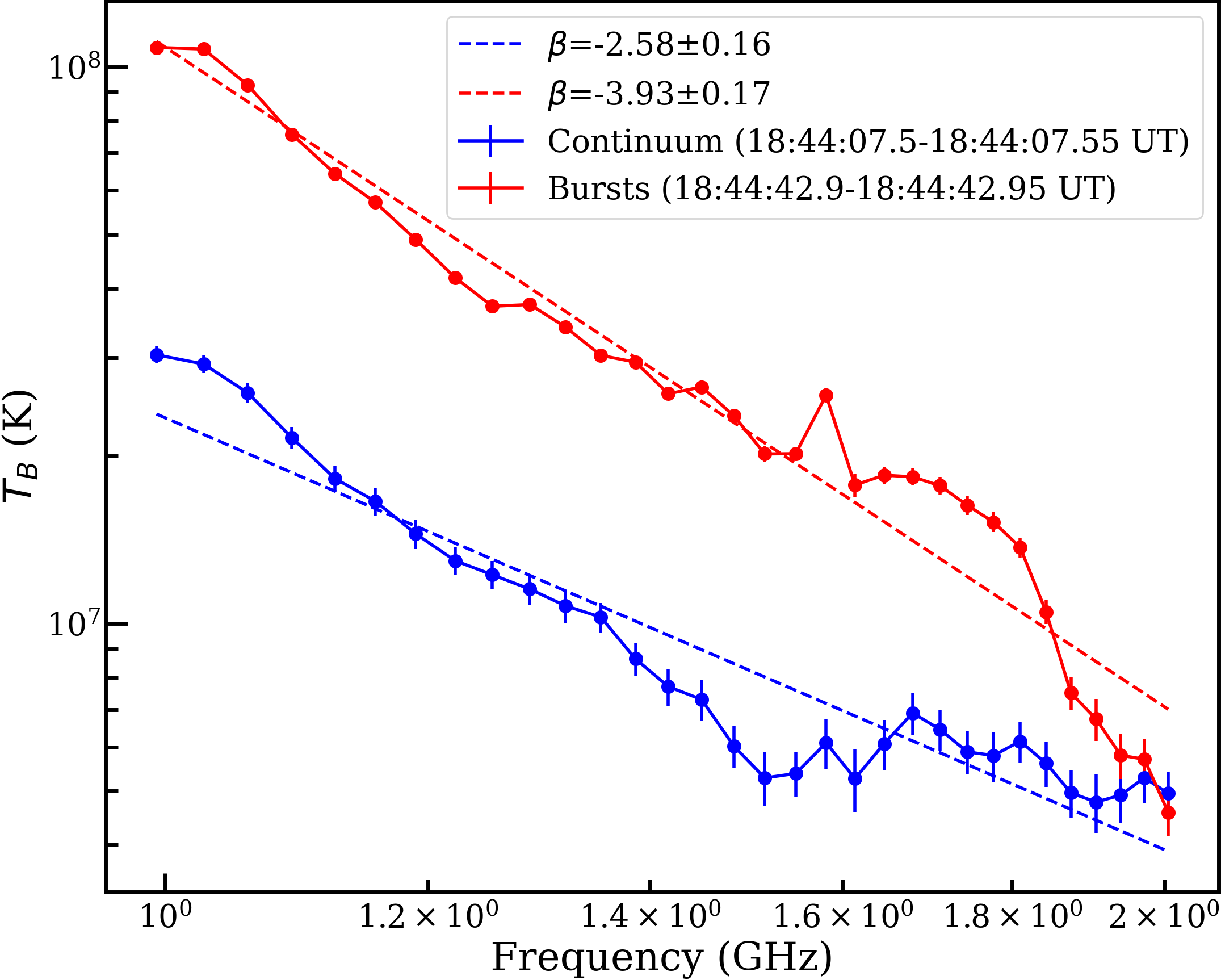}}  
         & \resizebox{55mm}{!}{
\includegraphics[trim={0.0cm 0cm 0.0cm 0.0cm},clip,scale=0.3]{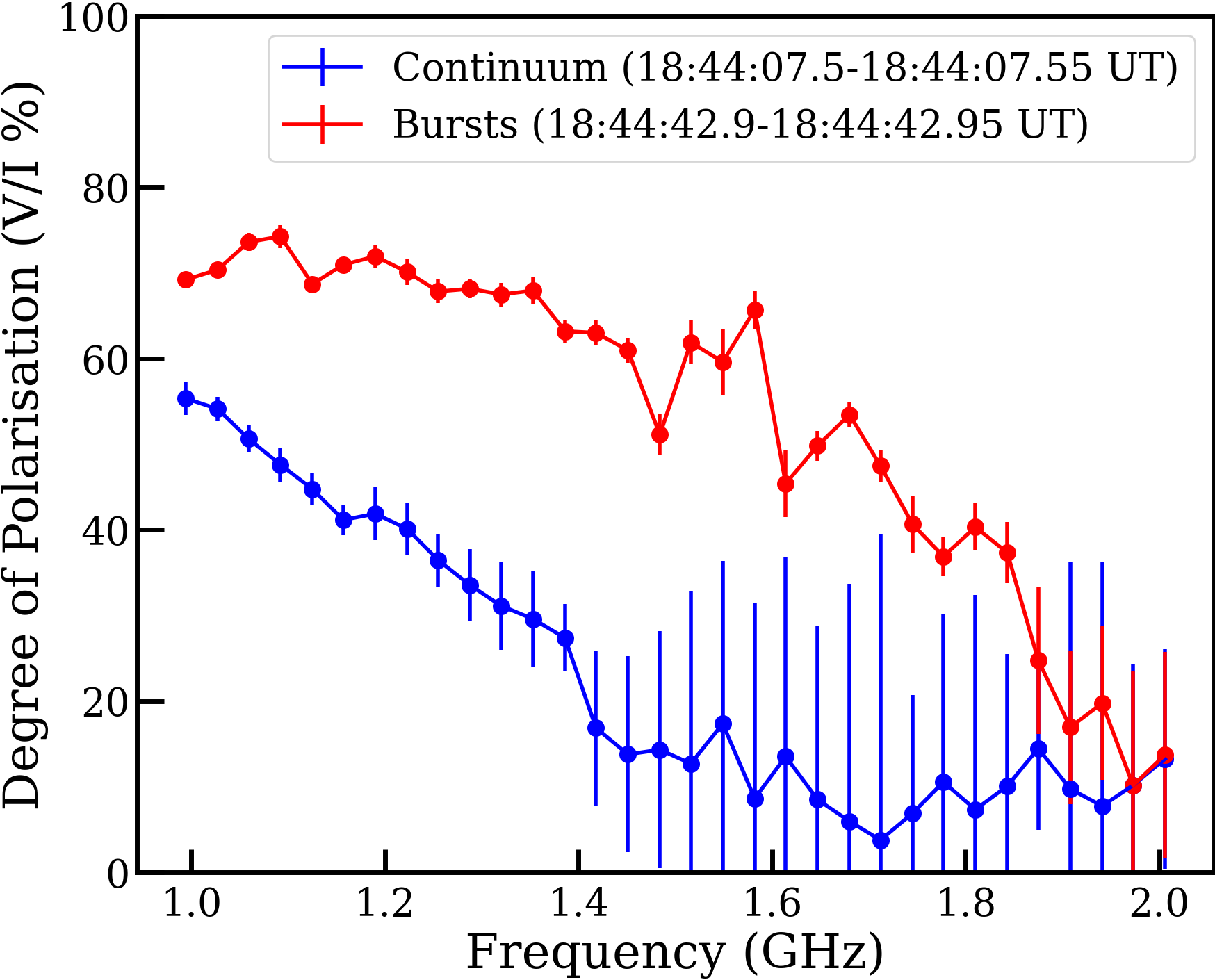}}  &
\resizebox{52mm}{!}{
\includegraphics[trim={0.0cm 0cm 0.0cm 0.0cm},clip,scale=0.3]{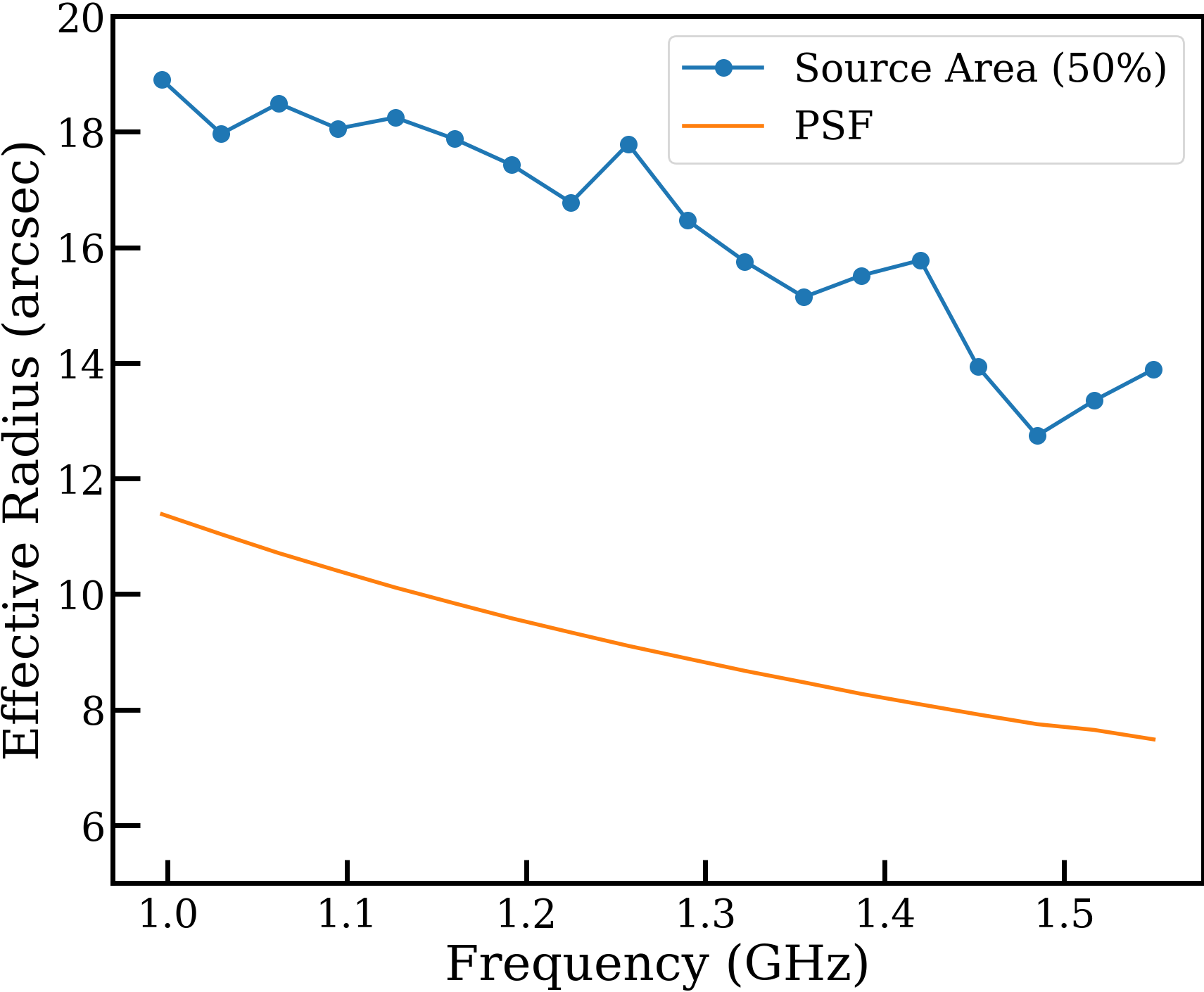}}  
\\
        (B) $T_B$ & (C) Degree of Circular Polarisation & (D) Area of the source
    \end{tabular}

\caption{Panel A: The maximum brightness temperature variation of the radio source at 1 GHz for the entire observation computed over the radio map. The yellow shaded region shows the continuum emission duration, while gray region shows radio burst duration. Panel B: Frequency spectrum of the maximum $T_B$ of the radio source from the images during the continuum emission and the bursts. The errorbars are the 1-$\sigma$ error, where $\sigma$ is the RMS from the region far away from the radio source. The blue and red dotted lines show the power-law fits (Sec. \ref{subsec:sst_temporal}) to the spectrum for continuum emission and bursts, respectively. Note that steep slope values are in the legend panel. Panel C: The degree of polarisation (ratio of stokes~V and stokes~I brightness temperature) of the radio source computed from the stokes-V images for continuum emission and bursts. The errorbars are computed via same process as in panel B. 
Panel D: The variation of the effective radius of the area at 50\% contour of the radio burst with frequency. The effective radius is $r_{eff} = \sqrt{A_{s}/\pi}$, where $A_{s}$ is the source area . The profile of the effective radius of PSF ($= \sqrt{b_{min}*b_{max}}$) with frequency is shown in orange, where $b_{min}$ and $b_{max}$ are minor and major axis respectively. Note that the effective area is shown for the peak of the radio burst, i.e. at 18:44:42.9 UT.  }
\label{fig:lineplots}
\end{figure*}

Figure \ref{fig:aia}(A) shows the flare location in the AIA 171 \AA \ map and radio source before the burst. We note that the location of the bright radio source is far away from the flare site ($\approx 75"$). The radio burst source occurs at the same location as a bright radio source before the bursts (Fig. \ref{fig:aia}(C)) at (-750",180"). When superimposed with the HMI magnetogram (Fig. \ref{fig:aia}(B)) and AIA 171 \AA \ (Fig. \ref{fig:aia} (C)), we can clearly observe the separation between the radio burst location and the flare location. The image also shows the footpoints of the large-scale loops near the sunspots (-750",180") and in the away regions (-850",200"). Figure \ref{fig:aia}(B) shows the radio burst source for different frequencies of L-band. We note that the source is compact at the high frequencies (e.g. 1.4 GHz) and marginally extended at 1 GHz in North-South direction in a kidney-bean shape. The extension at 1 GHz is roughly twice the beam size shown in blue ellipse. Fig. \ref{fig:lineplots} (D) shows that the source is resolved at 50\% contour levels w.r.t. synthetic beam. In addition, the center position of the high-frequency source is closer to the sunspot than low-frequencies (Fig. \ref{fig:aia} (B)).

\subsection{Spectral \& Temporal Features}
\label{subsec:sst_temporal}
Figure \ref{fig:lineplots}(A) shows the light curve of the radio bursts for 1 GHz for the maximum $T_B$ of each radio map. The brightness levels are much higher, with a brightness temperature of $\sim$ 20--30 MK, compared to the quiet Sun levels. We observe temporal variations in radio emission, and these bright features can be seen as a mixture of second and sub-second-level variations. The yellow region in Fig. \ref{fig:lineplots}(A) shows a wavy emission structure in time (e.g. 18:40 to 18:41 UT) with spiky emission of sub-second scale (e.g. 18:43:22 UT). Fig. \ref{fig:lineplots}(A) grey region shows the radio bursts to possess similar temporal variations with an additional seconds timescale oscillation between 18:44:30 UT to 18:45:14 UT. The radio bursts are characterized by a prominent rise of 40 MK above the continuum levels. The brightness temperatures of the radio bursts are higher by an order of magnitude than the continuum time. We see seven distinct periodic peaks between 18:44:40 UT to 18:45:00 UT in the radio burst as a QPP. The radio emission before and during this QPP shows a steep spectral dependence shown in Fig. \ref{fig:lineplots}(B), which shows the frequency spectrum of the radio source for the bursts (red curve) and the continuum source (blue curve), respectively. We note that a steep power-law index ($\beta$) of $-2.58$ and $-3.93$, where $T_B \propto \nu^{\beta}$ and $\nu$ is the frequency respectively. Both these components are also circularly polarised. Fig. \ref{fig:lineplots}(C) shows the degree of polarisation, defined as the ratio between stokes V and stokes I. The radio bursts show a nearly constant and high degree of right-handed circular polarization of $\approx 60-70\%$ at 1 GHz during continuum ECM and bursts. A high degree of circular polarisation is a characteristic of ECM emission \citep{Vlahos1987}. The circular polarisation decreases from 1.0 GHz to 1.5 GHz, reaching up to 40\% before the bursts at 1.2 GHz, and it decreases to almost ($<20$\%)) until 1.5 GHz. For bursts, the circular polarization goes to $<20$\% around 1.9 GHz. For frequencies $>$1.9 GHz, the circular polarisation of bursts is lost. This could be a result of the fainting of the intrinsic ECM source as we probe the higher frequencies and lower coronal heights.

\subsection{Conditions for ECM emission}
\label{subsec:model}

To study the origin of the radio source, its location w.r.t. ambient magnetic field topology is crucial. The radio emission mechanism from the energetic electrons depends on the wave-particle and wave-wave interactions, i.e. indirectly on the ambient magnetic fields and densities in the loops, and incoming electron energy distributions. We build a density and magnetic field model, which provides the framework to highlight the origin of the radio sources and the overall event picture.

\begin{figure}
    \centering
\resizebox{65mm}{!}{
\includegraphics[trim={0.0cm 0cm 0.0cm 0.0cm},clip,scale=0.6]{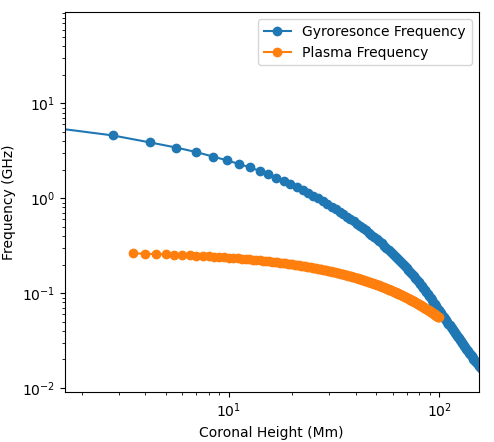}}
    \caption{1-D profiles of the plasma and gyro-resonance frequencies near the observed radio bursts location. Note that both gyroresonance frequency and plasma frequency equals at $\approx$ 100 Mm.}
    \label{fig:dens_model}
\end{figure}

\begin{figure*}
    \centering
    \begin{tabular}{cc}
    \resizebox{70mm}{!}{
\includegraphics[trim={0.0cm 0cm 0.0cm 0.0cm},clip,scale=0.6]{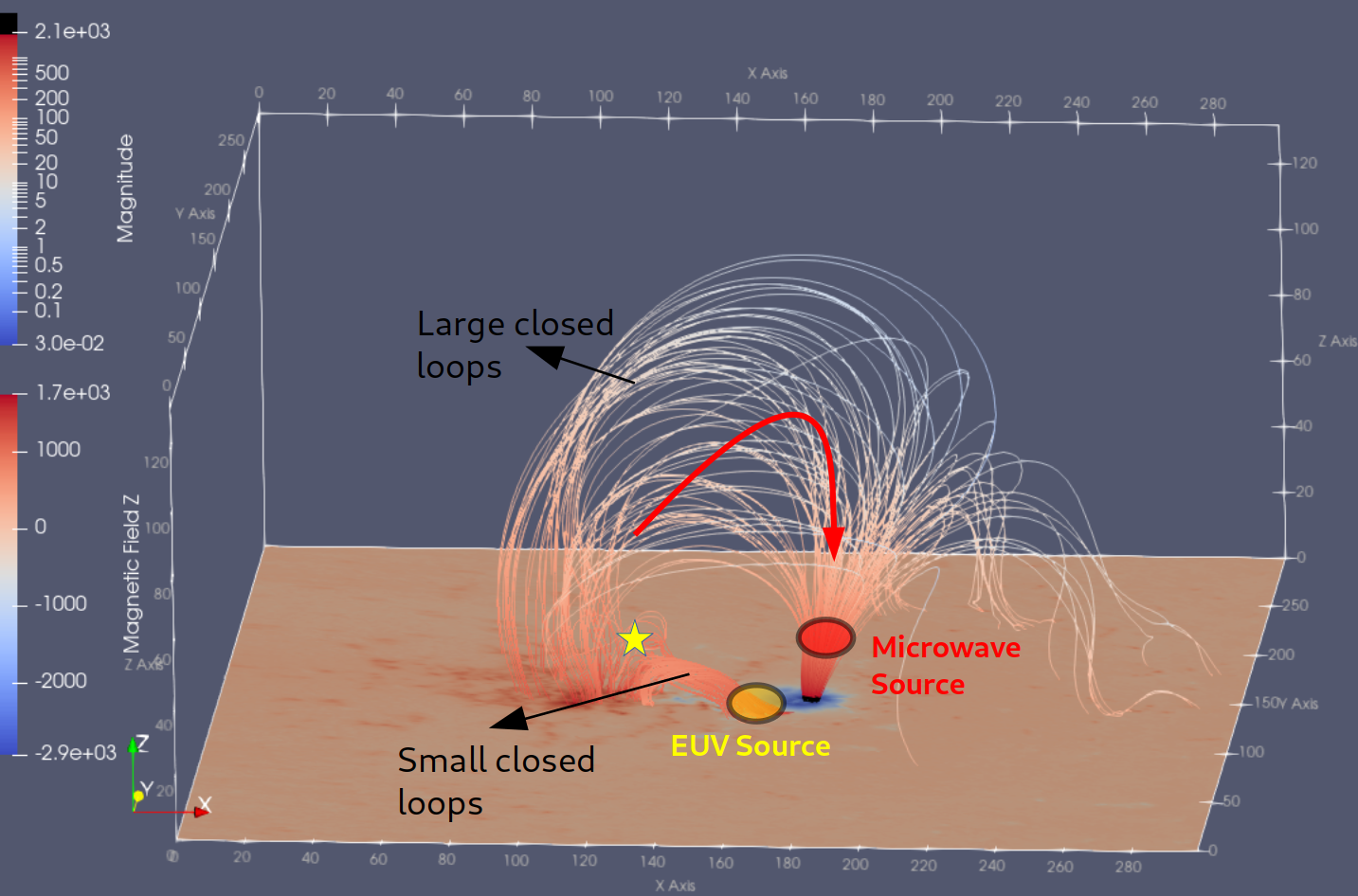}}     & \resizebox{80mm}{!}{
\includegraphics[trim={0.0cm 0cm 0.0cm 0.0cm},clip,scale=0.6]{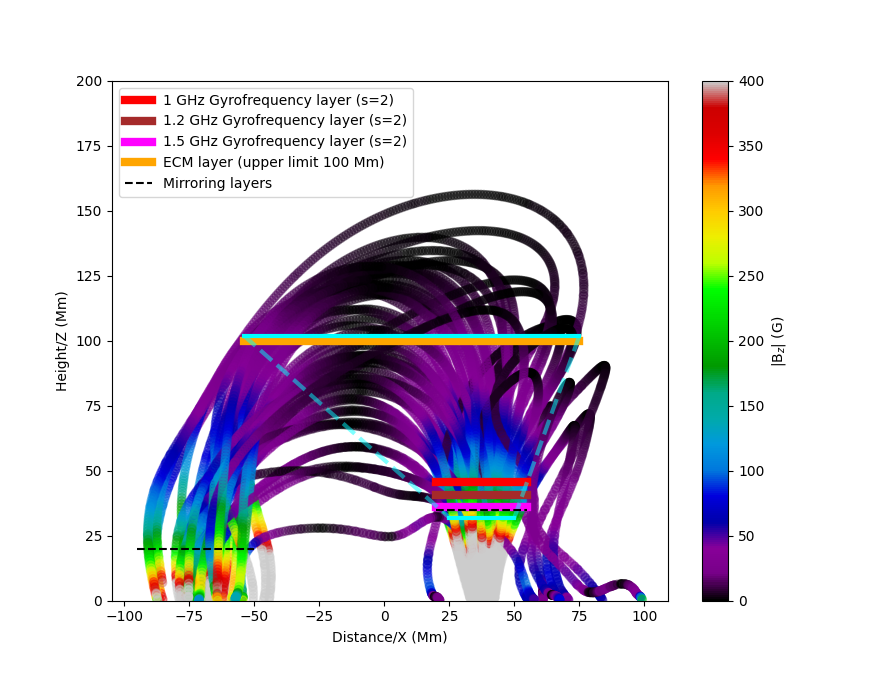}}  \\
    (A) Magnetic extrapolation model & (B) $|B_z|$ in X-Z plane \\
    \resizebox{70mm}{!}{
\includegraphics[trim={0.0cm 0cm 0.0cm 0.0cm},clip,scale=0.6]{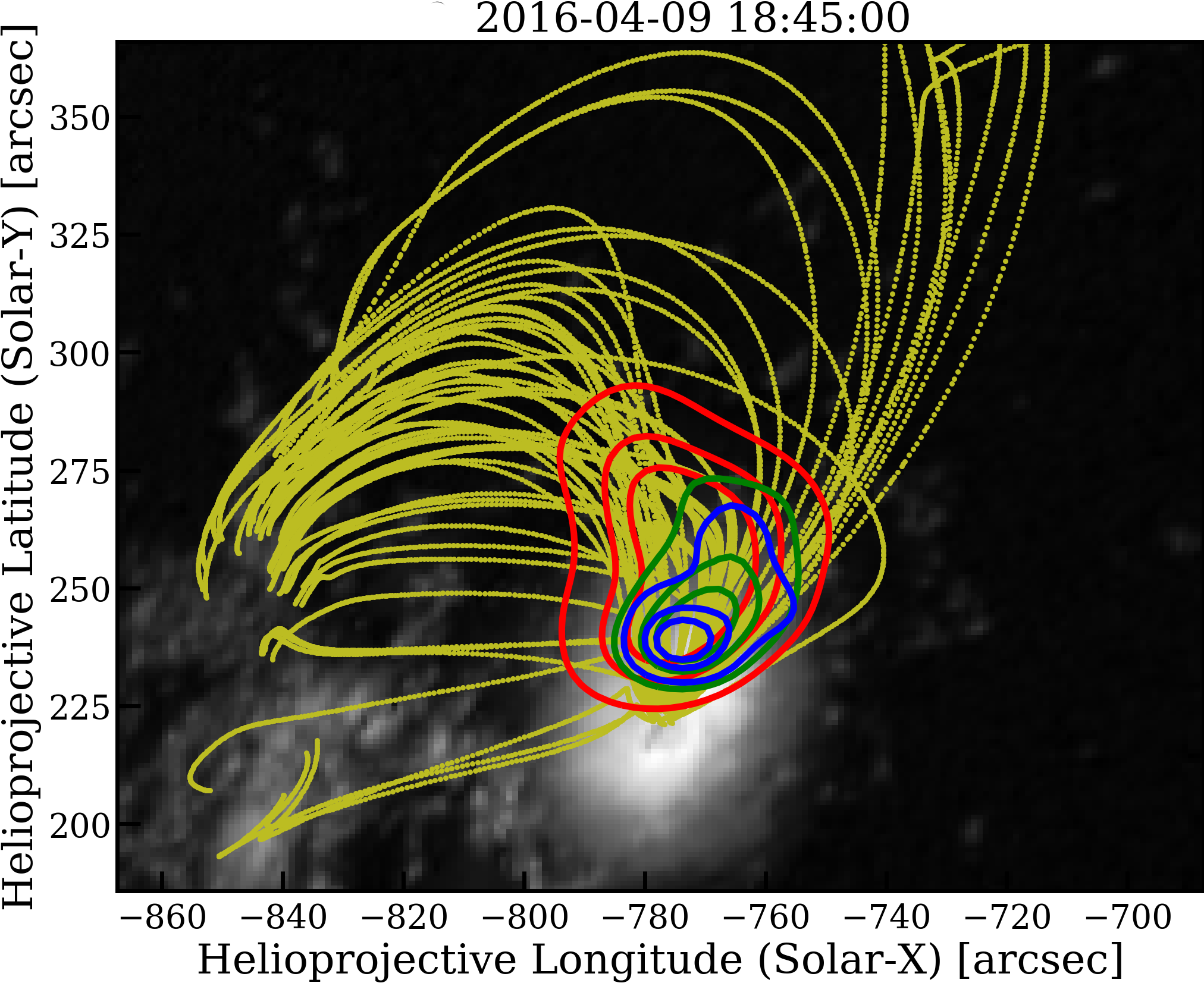}}   &  \resizebox{85mm}{!}{
\includegraphics[trim={0.0cm 0cm 0.0cm 0.0cm},clip,scale=0.6]{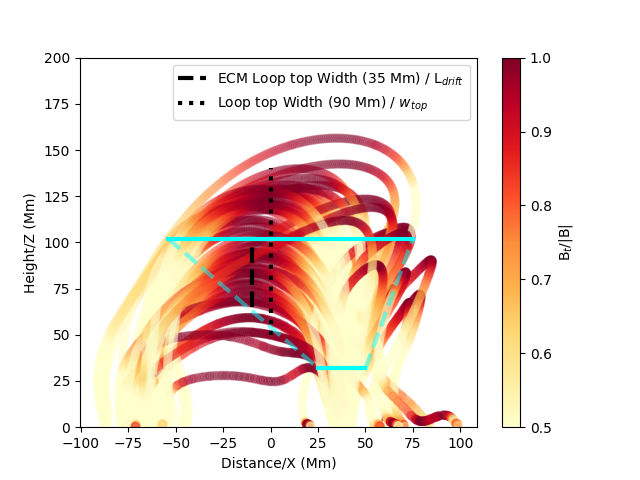}} \\
      (C) Projected magnetic fields   & (D) $|B_{t}|/|B|$ in X-Z plane \\
    \end{tabular}

    \caption{Panel A: Schematic of the 3-D magnetic extrapolation model showing the approximate location of the reconnection site, ECM radio source, EUV flare source and the coronal loops connecting the reconnection site to the observed radio source locations. Note that two sets of smaller and larger coronal loops are responsible for the electron transport producing EUV and radio sources, respectively. The yellow star marks a tentative location of the reconnection region near the confluence of the big and small magnetic loop. Panel B: X-Z plane projection of the large-scale magnetic loops colour-coded with the absolute magnitude of the $B_z$. The red bar shows the gyro-resonance layer for s=2 harmonic at 1 GHz, while the magenta bar is the same for 1.5 GHz. The mirror point is shown in a black dashed line. The cyan trapezoid marks the tentative ECM source region. Note that the dashed cyan lines are straight lines connecting the ends of 1.5 gyrofrequency layers with the uppermost possible ECM layer. Panel C: The extrapolated magnetic field re-projected on the HMI maps, along with the radio contours of the bursts. The blue, green and red are 1.5, 1.2 and 1.0 GHz respectively. Panel D: The ratio of the tangential magnetic field and total magnetic field in the X-Z plane. The cyan trapezoid marks the tentative ECM source region, the same as panel B. The vertical dotted line is the approximate width of the loops at the top. The vertical dashed line is the approximate ECM region width at the loop top and is discussed in detail in section \ref{sec:model_ecm}.}
    \label{fig:model}
\end{figure*}

To obtain the radial electron density profile at the radio burst site, we use PSIMAS ( Predictive Science incorporated Magnetohydrodynamic Algorithm outside a Sphere) model\footnote{https://www.predsci.com/corona/model\_desc.html} for 9 April 2016, 18:45:00 UT. PSIMAS comprehensively models magnetic fields, densities, and temperature in a realistic and self-consistent way using resistive 3-D MHD. The electron plasma density ($\omega_p$) at the site of the burst is plotted in Fig. \ref{fig:dens_model}.
From the PSIMAS 3-D density cube, we extract the 1-D radial profile near the radio burst site at (-780", 230"). We can approximate this 1-D density profile. We assume a cartesian 3-D geometry of XYZ, where X and Y are in the plane of the solar surface, while Z is the radial height direction. A typical form of a 1-D coronal model of electron density is an exponential,i.e.,
\begin{equation}
    n_e (Z) = n_1 e^{-Z/Z_1},
    \label{eq:den}
\end{equation}
where $n_e$ is the electron density profile, $n_1$ is the base coronal electron density, and $z_1$ is the coronal scale height, respectively. For the selected 1-D density profile, we get the values of $n_1\approx9.5\times10^8$~cm$^{-3}$ and $z_1\approx$~31 Mm using an exponential fit (Eq. \ref{eq:den}). 

Further, we perform a 3D coronal magnetic field extrapolation to explore the magnetic field topology and radio burst location. We use $gx\_simulator$ package (part of the IDL SolarSoftware distribution; \cite{Nita2015}) and make a 3-D magnetic field vector cube for the active region shown in Fig. \ref{fig:model} (A) using nonlinear force-free field (NLFFF) method. We show bigger loops connecting the site of the EUV brightening to the site of the radio burst over the sunspot. Small closed loops are the loops brightened in EUV 171 \AA\ during the flare. Therefore, the magnetic reconnection site is likely to be in between them, their meeting region close to the footprints. A tentative reconnection region near the confluence of the bigger and smaller magnetic fields is shown as a yellow star in Fig. \ref{fig:model} (A). The cross-sectional geometry of the loops in the X-Z plane is shown in panel B. Figure \ref{fig:model} (C) shows the magnetic fields projected on the HMI maps along with the radio contours for the radio bursts at 1.0, 1.2, and 1.5 GHz. We note that radio burst sources nicely match the projected large-scale loop in their location. The extrapolation yields a magnetic vector at each grid point in the simulation cube; hence, we construct the tangential component of the magnetic field ($B_{t}$) along the bigger loop w.r.t the local magnetic field vector. We note that the ratio of $B_{t}/|B|$ is almost unity at the loop top. For a qualitative calculation, we construct a 1-D radial magnetic field strength.
To build a 1-D magnetic field model near the observed radio burst location, we extract the radial profile from the 3-D magnetic field cube near the site of the radio burst. Fig. \ref{fig:dens_model} shows the gyro-frequency ($\Omega_B$) radial profile using relation $\Omega_B=2.8 B$(Gauss).

\begin{table}[h]
    \centering
    \begin{tabular}{|c|c|}
    \hline
        Parameter & Values\\
    \hline \hline
        Total loop length & $\approx$ 250 Mm \\
        Magnetic scale height ($\lambda$) & $\approx$ 80 Mm  \\
        (B$_{top}$, B$_{m}$) & (40, 200 G) \\
        X-ray-radio time delay ($t_{delay}$) & 38 s \\
        ($h_{LMP}$, $h_{RMP}$) & (20, 35 Mm)  \\
        ($h_{1GHz}$, $h_{1.2GHz}$, $h_{1.5GHz}$) & (46, 40, 36 Mm)  \\
        ($n_{e,top}$, $n_{e,1GHz}$, $n_{e,1.5GHz}$) & $(0.3, 1, 6)\times10^8$ cm$^{-3}$ \\
        Loss cone at top ($\alpha_{0}$) & 26$^o$ \\        
        Trapping length ($L_{trap}$) & 195 Mm \\
        Growth rate ($\Gamma$; o-mode;s=2)& 4.0 s$^{-1}$ \\
        Mean free path ($\lambda_{ei}$) & 0.7 Mm \\
        ECM source Area (A) at 1 GHz & 520 Mm$^{2}$ \\
        \hline
    \end{tabular}
    \caption{Table showing the values of various parameters obtained using density and magnetic field model. Note that the values are rounded to the next decimal place. The $h$ represents a coronal height from the solar surface. Note that the $h_{LMP}$ and $h_{RMP}$ are the estimated coronal heights of the eastern (left) and western (right) mirroring points.
    The $n_{e,top}$, $n_{e,1.5GHz}$ are the electron densities eastern (right) at the loop top and the gyrofrequency layer corresponding to 1.5 GHz emission, i.e. $h_{1.5GHz}$. The $n_{e,100Mm}$ is the electron density at the 100 Mm coronal height, i.e. start of the masing volume.}
    \label{tab:2}
\end{table}

\subsection{Observational ECM Source heights}
\label{subsec:ecm}
Figure \ref{fig:dens_model} shows the constructed gyro-frequency and plasma density radial profiles. We note that the magnitude of the gyro-frequency exceeds plasma frequency for a considerable range of coronal height, one of the essential conditions for ECM. The other emission mechanism candidate, i.e., plasma emission, is expected in denser and low magnetic field regions, which is unlikely over a sunspot. Therefore, the plasma emission mechanism is unlikely to dominate over the ECM emission \citep{Melrose1991}. \cite{Yu2023} also found that the plasma radiation can not explain the observed spatial distribution in frequency unless a $<$10 Mm density scale height is present above the sunspot.
For the coronal heights less than 100 Mm, the ratio $\omega_{pe}/\Omega_B <1$, making the region suitable for ECM instability (Fig. \ref{fig:dens_model}). In addition, the steep spectral behaviour observed in radio emission is unlikely to be from incoherent emission mechanisms like thermal bremsstrahlung and gyro-resonance. Overall, a persistently high degree of RCP, spatial location on the footpoint of the large-scale converging loop geometry at the sunspot, steep spectral power-law index, and $\omega_{pe}/\Omega_B<1$ coronal conditions support the evidence of the observed radio bursts is most likely to be from ECM source. From the magnetic extrapolation, the magnetic loop connecting the flare site and the sunspot is shown in Fig. \ref{fig:model} (A \& B). We note that the converging magnetic field over the sunspot at $X=25$--50 Mm is favourable for coherent ECM, while the other footpoint shows non-converging complex fields. \cite{Yu2023} already constrains the radio source emission to be $s=2$ o-mode from the opacity calculation for the same sunspot and shows the $s=2$ o-mode is most likely to escape. We observe the RCP radio emission over a negative magnetic polarity confirming an o-mode emission. Fig. \ref{fig:model} (B) also shows that the gyro-resonance layers for $s=2$ for 1 GHz, 1.2 GHz and 1.5 GHz are located at $Z_{1GHz}$ = 46 Mm, $Z_{1.2GHz}$ = 40 Mm and $Z_{1.5GHz}$ = 36 Mm respectively. All these heights are measured from the base of the right footpoint (Fig. \ref{fig:model} (B)) and shown in Table \ref{tab:2}. An upper limit on the ECM volume can be deduced where $\Omega_{B}<\omega_{r}$ at a height to $\approx$100 Mm (Fig. \ref{fig:dens_model}).  Combining the above calculations with the feasible coronal ECM volume, we find the extent of the possible ECM-unstable volume from $36<Z<100$ Mm, spanning $64$ Mm in coronal height. The X-Z plane cut of this possible ECM-unstable volume is shown as cyan trapezoid in Fig. \ref{fig:model} (B). Note that the upper limit of the ECM volume ($Z\approx100Mm$) partially spans the loop top region.

\section{Analysis of Time Delay in ECM}
\label{sec:model_ecm}
This section aims to characterize some properties of the electrons producing ECM radio source. Accelerated electrons with sufficient energy can travel to the ECM site over the sunspot separated by $\approx120$ Mm  (in X-Z plane) with an observed time delay ($t_{delay}\approx38$s). The delay depends predominantly on collisional and magnetic mirroring timescales.




\begin{table*}[]
    \centering
    \begin{tabular}{|c|c|c|c|c|c|c|}
    \hline
    Energy & $v_e$   & $L_{stop,top}$ /$L_{stop,1.5GHz}$ (Mm) & $\tau_{stop,top}$/$\tau_{stop,1.5GHz}$ (s) & $L_{drift}$ (Mm)\\
    \hline
1 keV & 19.0  Mm/s &  90.6  /  4.5 & 4.8  /  0.2 & 11.3 \\
2 keV & 27.0  Mm/s  & 362.3  /  18.1 & 13.5  /  0.7 & 22.5 \\
3 keV & 33.0  Mm/s  & 815.3  /  40.8 & 24.9  /  1.2 & 33.8 \\
\hline
4 keV & 38.0  Mm/s  & 1449.4  /  72.5 & \textbf{38.4  /  1.9 }& \textbf{45.0} \\
5 keV & 42.0  Mm/s  & 2264.7  /  113.2 & \textbf{53.8  /  2.7} & \textbf{56.3} \\
8 keV & 53.0  Mm/s  & 5797.6  /  289.9 & \textbf{109.3  /  5.5} & \textbf{90.1} \\
\hline
10 keV & 59.0  Mm/s  & 9058.7  /  452.9 & 153.2  /  7.7 & 112.6 \\
15 keV & 72.0  Mm/s & 20382.2  /  1019.1 & 283.6  /  14.2 & 168.9 \\
20 keV & 82.0  Mm/s & 36235.0  /  1811.7 & 439.7  /  22.0 & 225.2 \\
25 keV & 91.0  Mm/s  & 56617.2  /  2830.9 & 618.9  /  30.9 & 281.5 \\
30 keV & 100.0  Mm/s  & 81528.7  /  4076.4 & 819.4  /  41.0 & 337.8 \\   \hline
    \end{tabular}
    \caption{Various relevant electron propagation parameters for different electron energies. The stopping length ($L_{stop}$), and stopping time ($\tau_{stop} = \frac{L_{stop}}{v_{e}}$) for the loop top and at 1.5 GHz gyrofrequency layer are shown. The drift lengths ($L_{drift}$) due to diffusion at the loop top are shown in the last column. The range of $\tau_{stop}$  between loop top and 1 GHz gyrofrequency layer for the middle rows from 4 keV to 10 keV are consistent with the observed $t_{delay}$ ($\approx38$ s), and suitable $L_{drift}$ consistent with the loop top width $w_{top}$, and are shown in bold.}
    \label{tab:energy}
\end{table*}

\subsection{Electron Dynamics for the ECM}
\label{subsec:electron_dynamics}
Sustaining an ECM source for many hours would require an equilibrium between electrons responsible for the ECM instability via electron injection and losses (Fig. \ref{Fig:ds}). A persistent or intermittent supply of energetic electrons into the magnetic trap must be balanced by a continuous weak diffusion into the loss cone and the precipitation loss (Fig. \ref{fig:flowchart}). The leakage of the trapped electrons most likely happens at the reflections at the mirror points, which are dense and also due to turbulence in the loops. 

\subsubsection{Electron Injection}

Figure \ref{fig:model} (A) represents the magnetic field model showing the geometry of the magnetic fields and the connection for the accelerated electrons from the eastern section of the active region to the sunspot via the marked larger loops. The smaller loop shown is probably connected to EUV brightening seen as magenta spot in Fig. \ref{fig:aia}(B) at (-810", 205"). We note that the smaller loop's eastern leg mixes with the larger loop's eastern footpoint. Although given the limitations of the magnetic field modelling in resolution, force-free assumptions, etc., getting the exact field lines for the electron injection/propagation into the larger loop is highly non-trivial. We assume the energetic electrons are injected from the eastern footpoint region in the active region AR12529, which remains continually active before the solar flare and radio bursts. The electron injection into the larger loops can happen from this mixed loop region as this region hosts flux lines from both small and larger loops. An approximate position of the reconnection is marked as a yellow star in Fig. \ref{fig:model} (A) near the eastern bend close to the small loop top ($\approx$ 25 Mm high from the surface). For an electron with energy $E_{keV}$, the velocity ($v_e$) is given by,
\begin{equation}
    v_e (Mm/s) = 300\sqrt{(1-\frac{1}{(0.002 E_{keV} +1)^2})}.
    \label{eq:v}
\end{equation}
Table \ref{tab:energy} lists the relevant range of electron energies and velocities. 

\subsubsection{Electron Mirroring}
The converging magnetic field topology at the sunspot facilitates magnetic mirroring and trapping. The electrons can get captured in the large closed loops via trapping (Fig. \ref{fig:model} (A)). In the continuum phase, the active region AR1259, on average, emits at intermediate A- \& B-class level, i.e. the average energy of electrons in the trap would be $\leq 10$ keV \citep{Glesener2020}.
Under a simplistic model of a single flux tube, accelerated electrons with favourable pitch angle ($\alpha>\alpha_0$) will likely get trapped in the higher magnetic field arches, forming an electron reservoir. Here, the loss cone angle, $\alpha_0$ at the loop top is given by,
\begin{equation}
    \alpha_{0} = sin^{-1} (\sqrt{\frac{B_{top}}{B_{m}}}),
\end{equation}
where $B_{top}$ and $B_{m}$ are the magnetic fields at the loop top and mirror point, respectively. The coronal heights of the mirroring point are not trivial to estimate. However, we note that radio emission occurs predominantly below 1.5 GHz before the burst (Fig. \ref{Fig:ds} (B)). We note that during bursts, the FT features do occur until 1.9 GHz; however, overall, considering during and before QPP, the emitted radio power is mostly below 1.5 GHz as seen in Fig. \ref{Fig:ds}. Therefore, assuming a nominal 1.5 GHz ECM cut-off and corresponding second gyro-frequency layer as a nominal mirroring point for most electrons, we estimate $B_m\approx40$ G and $h_{RMP}\approx35$ Mm, where $h_{RMP}$ is the coronal height of the right (western) mirror point.
An approximate position of the mirroring layers is shown in Fig. \ref{fig:model} (B) and values tabulated in Tab. \ref{tab:2}. Using them, we get $\alpha_0 \approx 26^o$, i.e. the electrons can remain trapped from $\alpha>26^{\circ}$ and form an electron reservoir in the magnetic arch (Fig. \ref{fig:model} (A)), while the electrons with $\alpha<\alpha_0$ will precipitate into the sunspot. The left (eastern) leg and right (western) leg mirroring layer are at 20 Mm and 35 Mm from their respective loop base (Table \ref{tab:2}) for the 1.5 GHz gyrofrequency layer. The right leg over the sunspot shows a coherent variation of the $B_z$ magnetic field or gyrofrequency layers suitable for magnetic mirroring. The left leg is not coherent, i.e. the mirroring layer is just an average description of the mirroring points corresponding to varied field lines. Therefore, from the average 1.5 GHz gyro-frequency layers subtraction from the total loop length, an upper limit on the average, the trapping length ($L_{trap}$) is $\leq 195$ Mm. The trapping length for smaller electron energies will be shorter, and the mirroring points will be higher. The portion of this trapping region where columb collisions are strong, i.e. dense regions, will most plausibly make the electrons undergo into loss cone. Therefore, such denser regions near the loop leg over the sunspot will be suitable for ECM emission. We call this region as masing volume. The cyan curve in Fig. \ref{fig:model} (B) shows the extent of the masing volume, where the upper height cut-off comes from $\omega_p/\Omega_{B}<1$ limit, while the lower cut-off comes from the 1.5 GHz gyro frequency layer.

\subsubsection{Electron Reservoir}

The accelerated electrons responsible for the ECM must enter the loss cone within the masing volume via collisions. The extremely low electron energies would be stopped quickly due to collisions, while extremely high energies would be trapped for longer than observed $t_{trap}$ (38 s).
The stopping length ($L_{stop}$), i.e., the length travelled by an electron before stopping, can approximate the length travelled by the electron before entering the loss cone. Based on the density profile and a range of electron energies, we compute stopping distance assuming a thick target medium. 
\begin{equation}
    L_{stop} = \frac{E^2}{2Kn_e}
    \label{eq:Lstop}
\end{equation}
where, $E$ is the electron energy, $K=2\pi e^4 ln \Lambda$, $e$ is the electron charge \citep{Brown2002} and $ln\Lambda\approx20$. Corresponding to the $L_{stop}$, the time taken by the electron to stop will be $\tau_{stop} = \frac{L_{stop}}{v_e}$. The ambient thermal density governs the $L_{stop}$.
Table \ref{tab:energy} lists the $L_{stop}$ for the loop top and 1.5 GHz gyrofrequency layer. The loop top is the least, and the 1.5 GHz gyrofrequency layer is the most dense region of the masing volume. The 1.5 GHz is a high-frequency cut-off for the ECM, i.e., the stopping time of the electrons would be between the range of $\tau_{stop,top}$ and $\tau_{stop,1.5GHz}$. 

The coulomb collision and the magnetic fluctuations or turbulence can facilitate the entry of electrons into the loss cone. This diffusion of electrons is perpendicular transport across the magnetic field. We define $B_{\perp}$ as the turbulent perpendicular magnetic field component. We assume the maximum diffusion at the loop top, where $|B_{\perp}|/|B|\approx1$ \citep{Kontar2011}.
The perpendicular transport of fieldlines is given by $L_{drift}$ \citep{Kontar2011},
\begin{equation}
    L_{drift}  = \sqrt{2D_{M}L_{stop}},
\end{equation}
where $D_M = (\frac{B_{\perp}^2}{|B|^2} \lambda_{||}$ is the diffusion coefficient, $\lambda_{||}$ is the parallel correlation length of the perturbations. We assume the correlation length is of a similar order to the mean-free path (i.e. $\lambda_{||}\approx\lambda_{ei}$). The $L_{drift}$ must be constrained within the loop width of the larger loop, i.e. ($L_{drift}<w_{top}$) for coherent transport to the sunspot. Fig. \ref{fig:model} shows that the loop top region width is quite wide $w_{top}\approx90$ Mm, while it's $\approx35$ Mm falls in the masing volume.
Table \ref{tab:energy} lists the values of $L_{drift}$ for various electron energies. We note that the range of electron energies satisfying the time delay and perpendicular drift, i.e. $\tau_{stop,1.5GHz}<38\ s<\tau_{stop,top}$ and $L_{drift}<90$ Mm are 4-8 keV.

 \section{Analysis of QPP in ECM}
\label{sec:pulsations}

The VLA lightcurve shows QPP during the radio burst (Fig. \ref{Fig:ds} (C)). We distinguish the temporal variations in the radio bursts into two based on timescales, i.e. second level $>$ 1 sec and sub-second $<0.1$ sec timescales. The periodic variations will fall under the second timescale, while the latter is studied in the next section \ref{sec:fine}. 
We employ the running median subtraction technique to remove variations in desired timescales. In this approach, a continuum time series was formed with a smoothing time window of 20 sec, formed by calculating the median over it. Then, we subtract it from the main time series to get the desired running median subtracted time series. For this analysis, we chose the maximum $T_B$ time series from the radio maps. The maximum $T_B$ captures the variability of the radio source more robust than spatially-integrated flux density. Here, the choice with the smoothing window determines the variability retained in the subtracted time series. The continuum obtained for the former highlights the second-level undulations of the periodic pulses. This running median subtracted time series was passed to the wavelet analysis to characterise the varying second timescales further. An example of the running median subtracted time series with 20 sec smoothing window is shown in Fig. \ref{fig:wavelet} (A) top panel for 1 GHz made from $T_B$ time series shown in Fig. \ref{fig:lineplots} (A).

\subsection{Wavelet Analysis}

\begin{figure*}[h]
\begin{center}
\begin{tabular}{cc}
\resizebox{100mm}{!}{
\includegraphics[trim={0.0cm 0cm 0.0cm 0.0cm},clip,scale=0.3]{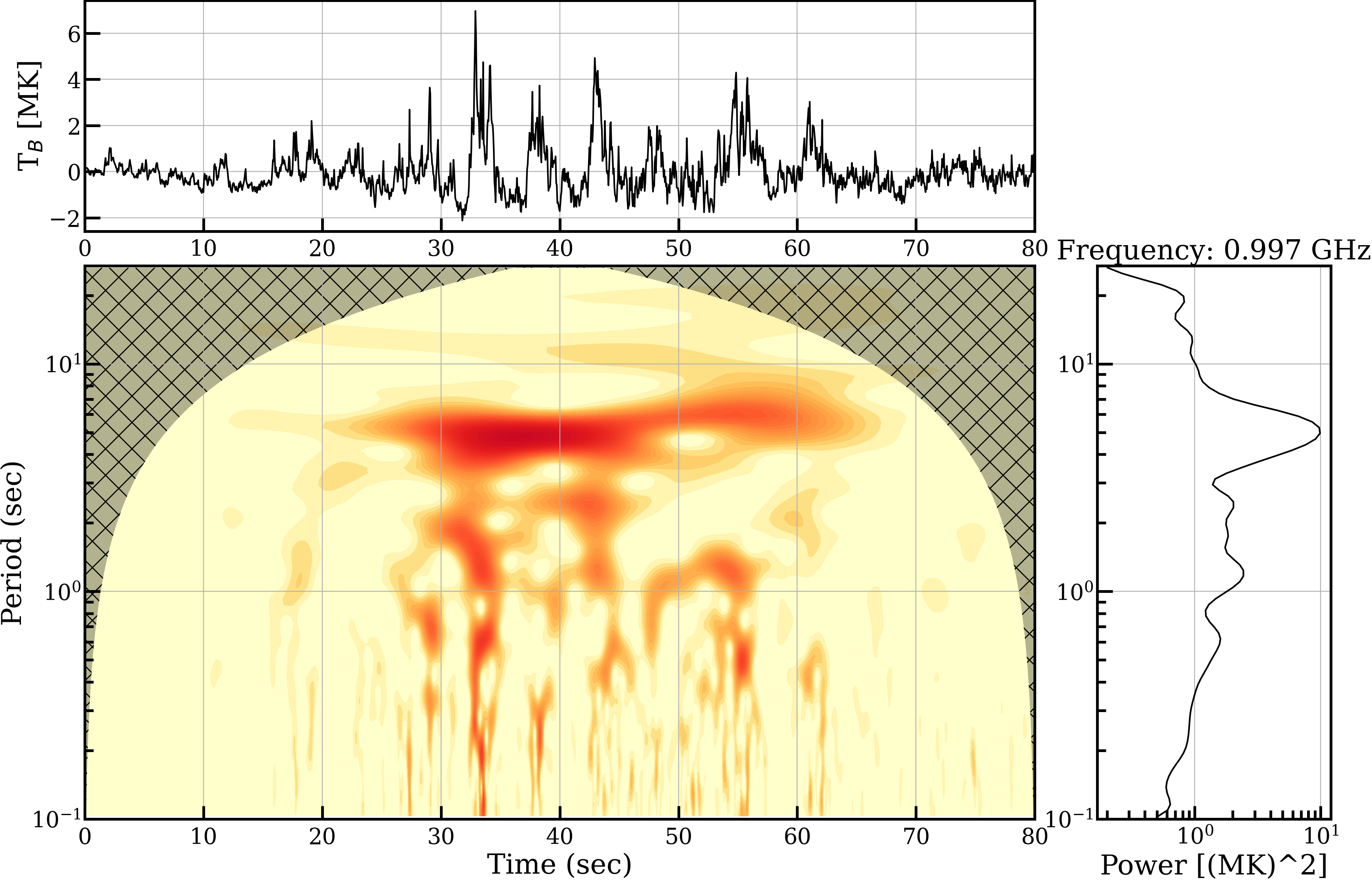}} 
&
\resizebox{80mm}{!}{
\includegraphics[trim={0.0cm 0cm 0.0cm 0.0cm},clip,scale=0.6]{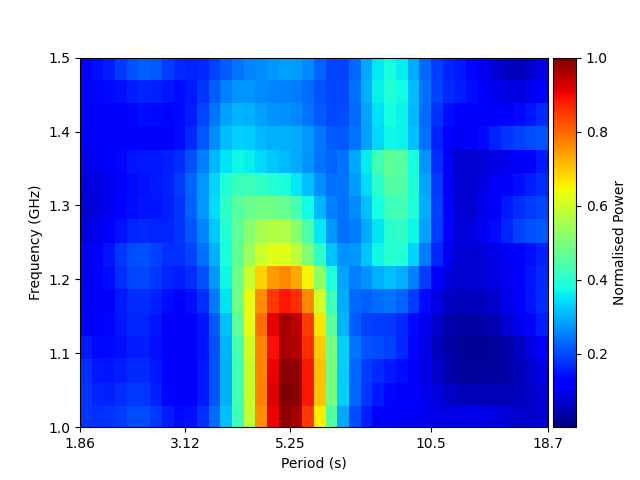}} \\
(A) QPP at 1 GHz & (B) Global Power Spectrum
\\
     \resizebox{86mm}{!}{
\includegraphics[trim={0.0cm 0cm 0.0cm 0.0cm},clip,scale=0.6]{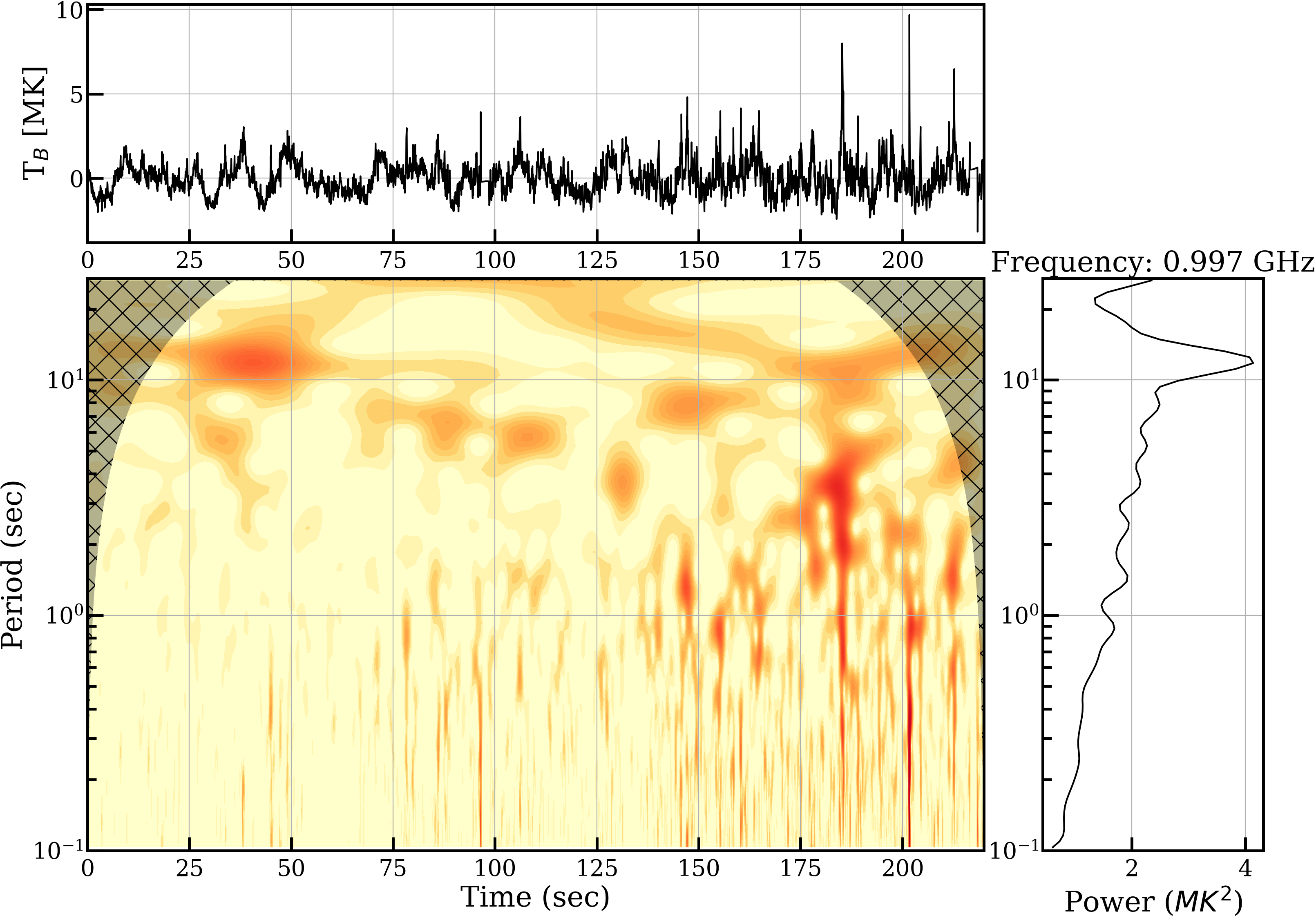}} 
     &  
     \resizebox{80mm}{!}{
\includegraphics[trim={0.0cm 0cm 0.0cm 0.0cm},clip,scale=0.6]{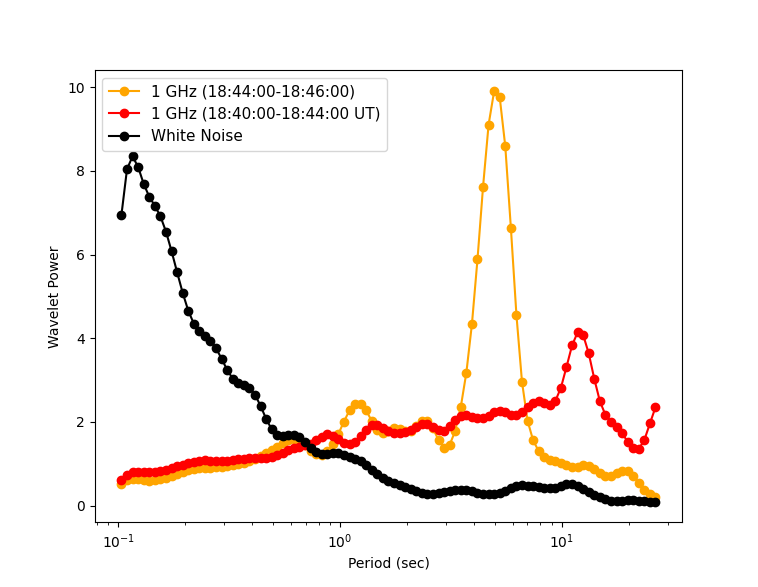}} \\
(C) Emission during non-QPP duration at 1 GHz & (D) Global Wavelet Power
\end{tabular} \\
\end{center}
\caption{Example of the wavelet analysis for the light curve at the frequency at 1 GHz. Panel (A): Wavelet power spectrum for 1 GHz time series for radio bursts. The top time-series is the running median subtracted maximum $T_B$ of the radio source. The start time is 18:44:10 UT. Panel (B): Normalised time-averaged wavelet power spectrum for each frequency channel from 1 GHz to 1.5 GHz. Panel (C): Wavelet power spectrum for 1 GHz time-series for the running median subtracted maximum $T_B$ before the QPPs. The start time is 18:40:10 UT. Panel (D): Line plot of global wavelet power spectrum for QPPs (orange) and non-QPP time (red). The black curve is the wavelet power for a simulated white noise time series.}
\label{fig:wavelet}
\end{figure*}

We perform wavelet analysis for all $T_B$ time series of 32 frequency bands in L-band for both the continuum and radio burst period. 
We chose the ``morlet" mother wavelet and temporal wavenumber corresponding to the timescales spanning 20 sec to 1 sec. 
Figure \ref{fig:wavelet} (A) shows a wavelet spectrum for the brightest 1 GHz time-series. During radio burst times, we clearly note a periodicity of 5 seconds. The right panel shows the time-averaged wavelet power with scales. 
Figure \ref{fig:wavelet} (B) shows the 2-D array for time-averaged wavelet power for each of 32 frequency channels. Here the entire array is normalized to unity. We note that wavelet power is at its maximum at a 5-second periodicity for the frequency range from 1.0 GHz to 1.3 GHz. 

Prior to the radio bursts, the radio continuum shows fine structure at a brightness level of 20-25 MK. The detailed study by \cite{Yu2023} shows a persistent sunspot microwave source lasting for many days, which is attributed to the sunspot "auroral" emission. Our present observation explores smaller time scales, i.e. minutes, seconds, and sub-second variations. We subject the continuum time series to wavelet analysis to find potentially weaker periodicities. Figure \ref{fig:wavelet} (C) shows the wavelet power spectrum for the 1 GHz.  We do not see any clear indications of periodicities before QPPs. The 5-sec periodicity of the radio bursts loses significance above 1.3 GHz (Figure \ref{fig:wavelet} (B)). The continuum emission global power does not show 5-sec periodicities but enhances uniform power above 1 sec. For a qualitative comparison, we produce a white noise time series and subject it to wavelet transformation. The data points for the white noise time series are derived from a zero mean normal distribution.
The continuum power is spread across the higher period compared to a typical white noise time series shown in black in Fig. \ref{fig:wavelet} (D).

\subsection{Saturation of ECM}
\label{subsec:dis2}
The brightness of the ECM emission depends on the positive gradient in an electron's velocity distribution function in the loss cone and the growth rate of the o-mode ($\Gamma\approx10^3$ $s^{-1}$). The ECM can be saturated in the loss cone due to a stagnant growth rate. This can occur if the scattering of electrons into the loss cone is limited or the loss cone is saturated by a finite rate at which it becomes empty. The latter mechanism dominates when $\delta=v_e/(L_{trap} \Gamma) < 1$ \citep{Melrose1982}.  We use typical masing volume estimates and most probable electron energies at 4 keV, i.e. $v_e$, $L_{trap}$ and $\Gamma$ are electron velocity, trapping length and growth rate, i.e. $42$ Mm/s, $195$ Mm and $10^3$ respectively. Therefore, we get $\delta=2.1\times10^{-4}$. We note that this scenario is feasible, and the finite emptying rate of the loss cone should determine saturation. The saturation level is reached in $\approx$2.5 sec, half of 5-second oscillations. At saturation, most of the free energy available gets converted into radiation and the saturated energy density \citep[$W$,][]{Melrose1982} for small pitch angles is given by,
\begin{equation} 
    W_{sat} = (\delta) n_e m v_e^2 \alpha_0^3.
\end{equation}
Here, the factor $\delta$ reduces the energy density saturation due to the finite discharge rate of the loss cone. We use $\alpha_0^3=3.7\times10^{-3}$ and the timescale of loss cone emptying is $\approx \frac{1}{\delta} = 1.5\times10^{3}$ sec. We also assume beam density $n_e\approx10^7$ cm$^{-3}$ (assuming X-ray nonthermal densities), and get $W_{sat}$ = $1.3\times10^{-7}$ erg cm$^{-3}$. For a masing volume of $V=A^{3/2}$ ($=1.18\times 10^{28}$ cm$^{3}$), we get volumetric radiation energy as $W_{sat}V\approx 2.0\times10^{21}$ erg. Further, the $T_B$ of a saturated ECM is approximated as \citep{Melrose1982,Yu2023},
\begin{equation}
   T_{B,sat} (K) \approx 2\times10^{12} (\frac{E}{1 keV})(\frac{\nu}{200 MHz})^{-2} (\frac{L_{trap}}{R_{\odot}})^{-1}.
\end{equation}
At 1 GHz frequency and 4 keV electrons, we estimate the $T_{B,sat}\approx5\times10^{11}$ K. This estimate is much higher than the observed maximum $T_{B,obs}\approx10^8$ K (assuming a synthesised beam size of 12$''$). Such contrast can be due to ECM operating at lower efficiency in an unsaturated regime and/or absorption in 3rd gyroresonance layer with optical depth($\tau_{0}$). Assuming simple radiation transfer, a back calculation of $\tau_0 =-ln(\frac{T_{B,obs}}{T_{B,sat}})$ gives $\tau_{0}\approx8.5$ ($<10$ necessary for transparent s=3 layer). Therefore, the opacity falls within a reasonable range estimated in \cite{Yu2023}.

\subsection{ECM Pulsations}
As found in previous sections, the geometry and electron propagation in the large loops are favourable for continuous operations of the ECM from the electron reservoir at the loop top. During the solar flare, additional acceleration of the electron population contributes to the reservoir and perturbs the continuous ECM operation. 
Unlike the continuum ECM phase (before 18:40 UT), the ECM burst shows a relatively spiky light curve and six quasi-periods of $\approx$ 5 seconds each (Fig. \ref{fig:wavelet}).  
Interpretation, like MHD-driven oscillation or sausage modes, does produce second timescale oscillations. However, it would result in a much smoother variation in brightness \citep[e.g.][]{Carley2019N},i.e. less feasible for observed spiky variation. In addition, the absence of 5 seconds during continuum emission also hints towards the lack of a persistent MHD waves-driven ECM source.


\cite{Aschwanden1988II} applied the Lotka-Volttera system to the spectroscopic observation of the QPPs. Here, the electron distribution and photon population naturally exchange energies, producing oscillations. This model is most feasible for our observation, where the incoming electrons cause perturbation to the distribution function and turn ECM into an oscillatory mode due to the particle and wave coupling. The period of such oscillation of the ECM emission is given by $\tau_p = 2\pi \sqrt{(\tau_{diff}\times \tau_{growth})}$, where $\tau_{diff}$ is the diffusion timescale of momentum of electron distribution, and $\tau_{growth}$ is the ECM instability growth timescale.
\cite{Aschwanden1990} estimates typically $\tau_{diff}$=5-20$\tau_{growth}$ using 2-D PIC simulation for o-mode ECM emission.
We use a nominal growth rate estimate $\tau_{diff}\approx10\tau_{growth}$. Therefore, using the observed pulsations of 5 sec, we get $\tau_{growth} \approx$ 0.25 s and $\tau_{diff}\approx$ 2.5 s.
We note that the diffusion timescales are larger than the trapping timescales ($\tau_{stop}$) mentioned in table \ref{tab:2}, especially at low energies. Large $\tau_{diff}$ implies ``weak" diffusion into the loss cone ($\approx10$ mins), which is a suitable mechanism to sustain continuum ECM sources given the availability of the energetic electrons. 

\section{Analysis of Sub-Second Fine Structures in ECM}
\label{sec:fine}
In this section, we focus on characterising sub-second structures in the ECM source. We utilize VLA's fine temporal resolution (50 ms) to study sub-second temporal variation present in the radio bursts. We use the same running median analysis employed in the previous section \ref{sec:pulsations}. Here, we chose a 20-second time window for studying second-level variabilities. 
Here, we characterise the shape of features in the frequency-time plane and compute power spectral density and wait-time distributions.

\begin{figure*}
    \centering
\resizebox{180mm}{!}{
\includegraphics[trim={0.0cm 0cm 0.0cm 0.0cm},clip,scale=0.6]{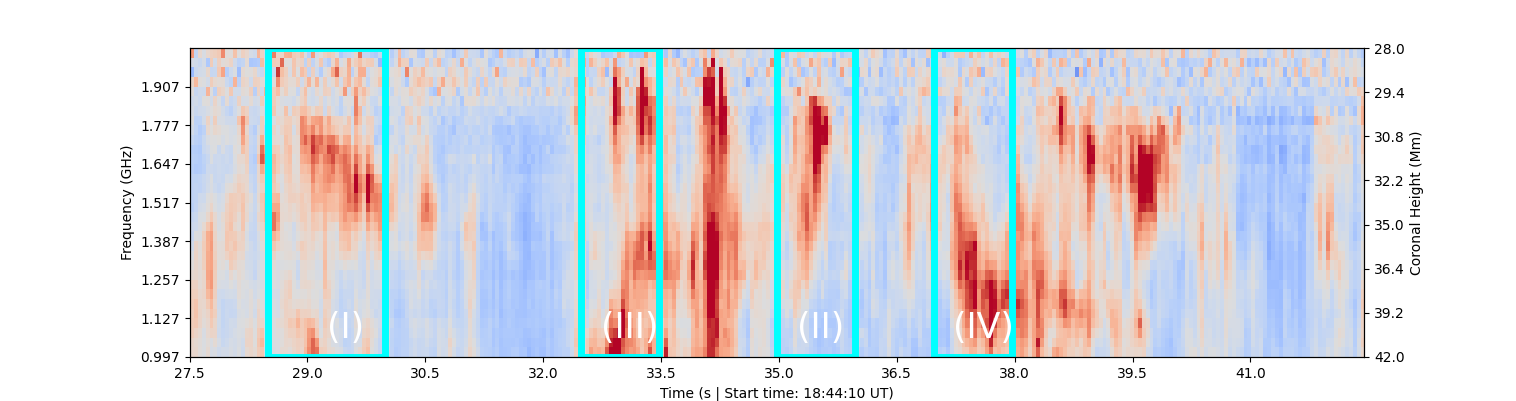}} \\
(A) FT Structures\\
    \begin{tabular}{cc}
   \resizebox{85mm}{!}{
\includegraphics[trim={0.0cm 0cm 0.0cm 0.0cm},clip,scale=0.6]{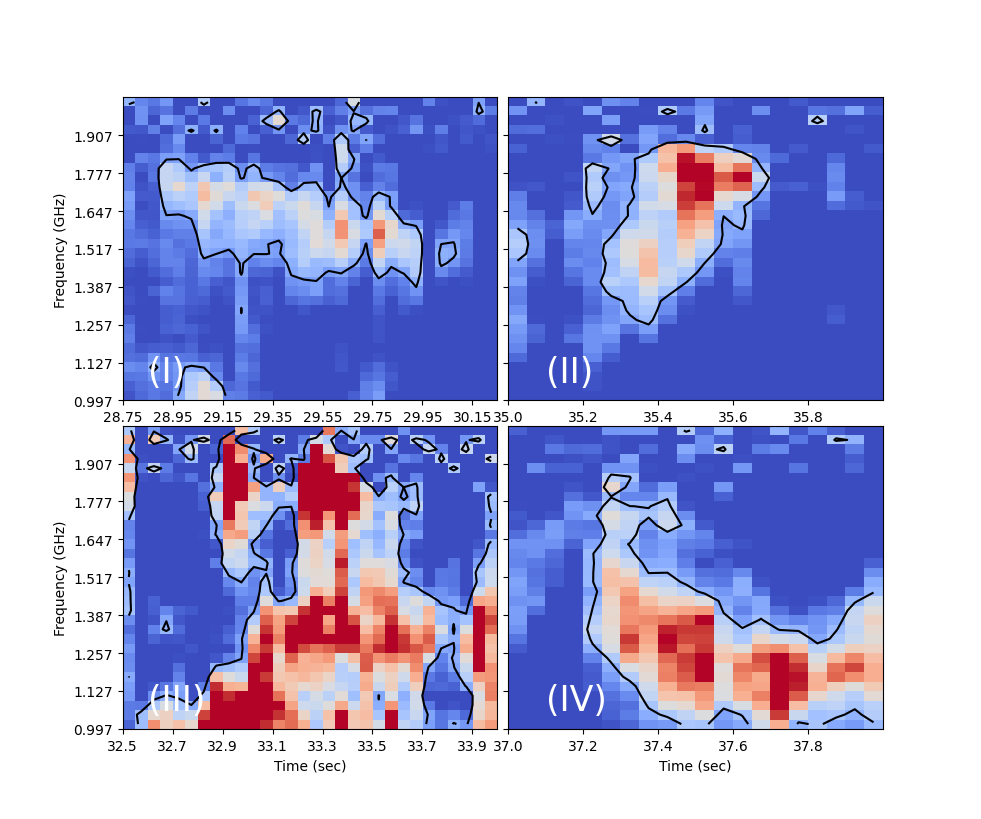}} &
   \resizebox{75mm}{!}{
\includegraphics[trim={0.0cm 0cm 0.0cm 0.0cm},clip,scale=0.6]{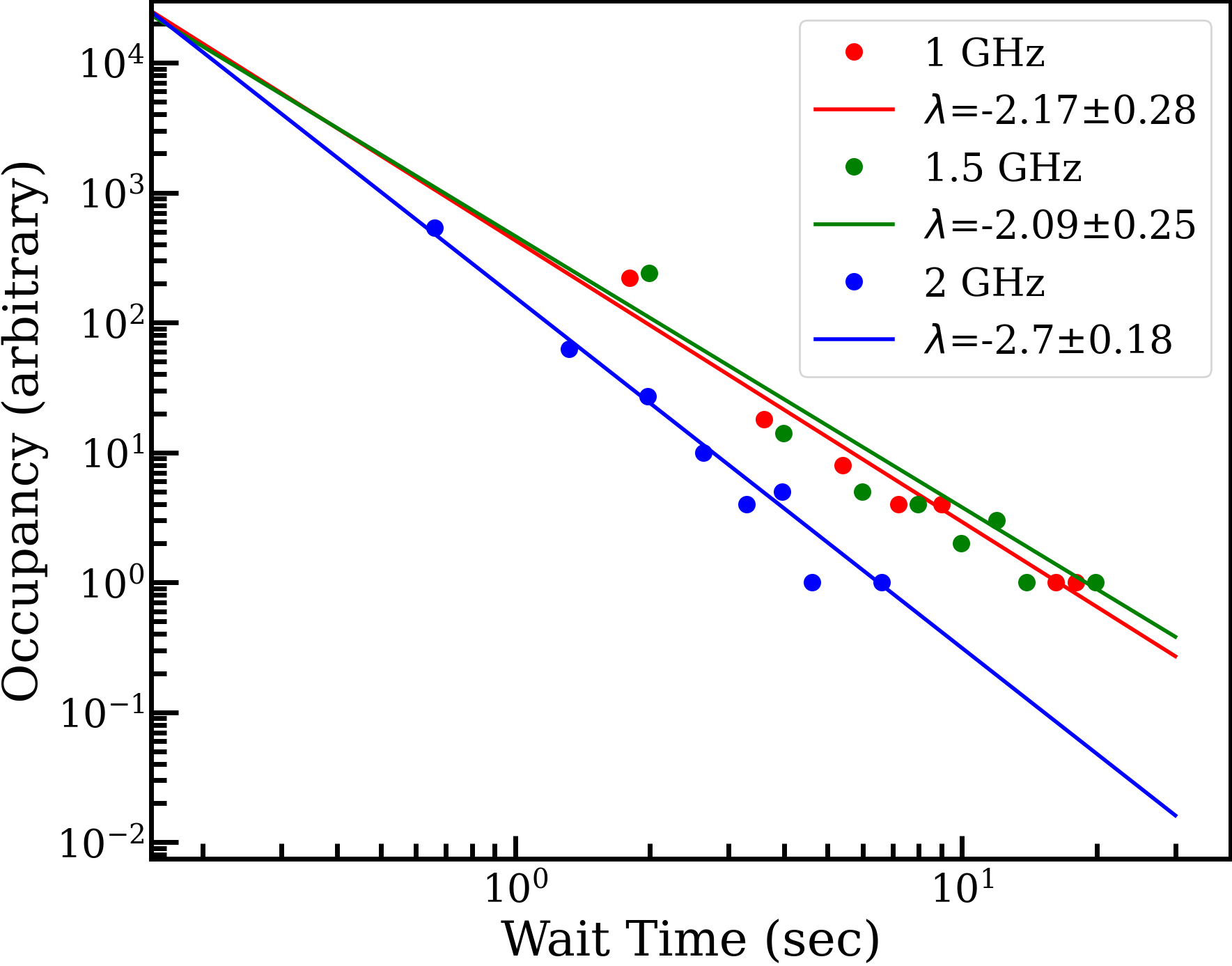}}\\
(B) Zoomed FT structures & (C) Wait-time distribution
    \end{tabular}
    \caption{Panel (A): A 15-sec segment of the running median subtracted DS showing the FT structures during the burst, with a 20-second smoothing window. Some individual coherent features are shown in sections I to IV. The section III is the peak of QPP. The start time, i.e. t=0, is 18:44:10 UT. Panel B: Zoomed-in DS is apparently showing drifting features for sections I to IV in panel A. Both positive and negative drift slopes are seen. Panel C: The data points correspond to the wait-time distribution of the time series of the fine structures for three frequency bands. The lines correspond to the linear fits done on each frequency band. Note that the slope for 1 GHz is slightly more gentle than the high-frequency slope.
    }
    \label{fig:sos}
\end{figure*}

\subsection{Frequency-Time Structures}

Despite the spectral-temporal behaviour's complexity, we observe some coherent Frequency-Time (FT) structures in the DS. These coherent structures occur as collective shapes in the FT plane and appear ``drifts". The FT structures are seen in the running median subtracted time series with a smoothing window of 20 sec. Figure \ref{fig:sos} (A \& B) shows four examples of the FT structures seen during a 15-second time segment. Note that the radio burst peak at 33.35 sec in Fig. \ref{fig:sos} (A \& B).  A zoomed-in version of these four FT structures (I-IV) is shown in Fig. \ref{fig:sos}. The three structures (II, III and IV) occur during the periodic bursts. Considering electron propagation, the flare 13 keV originated electron streams (Tab. \ref{tab:2}) and loop model would correspond to a frequency drift range of 1.85 GHz/s to 4.2 GHz/sec. Such a large drift rate will correspond to $\approx$ 100 MHz/50ms to $\approx$ 210 MHz/50ms. These drifts are much larger than the observed FT structure during the first pulsation, which has a 0.9 GHz/sec drift rate and is many times slower. Therefore, the FT structures seen are not produced by a single bunch undergoing instability; more likely a collective effect of many-electron streams undergoing instability. Assuming a characteristic Alfven wave speed at 1.0 to 1.5 GHz gyrofrequency ($\approx16-18$ Mm/s), the observed modulation drift rates can also have a contribution due to a fast-moving MHD disturbance. However, the exact dependence of FT structures on MHD waves is beyond the scope of this study. Here, the narrow bandwidth of the radio emission ($\sim 50$ MHz) suggests a quick dissipation of the electron beam within the ECM source with dissipation distance.

\begin{figure}
    \centering
    \begin{tabular}{c}
    \resizebox{80mm}{!}{
\includegraphics[trim={0.0cm 0cm 0.0cm 0.0cm},clip,scale=0.6]{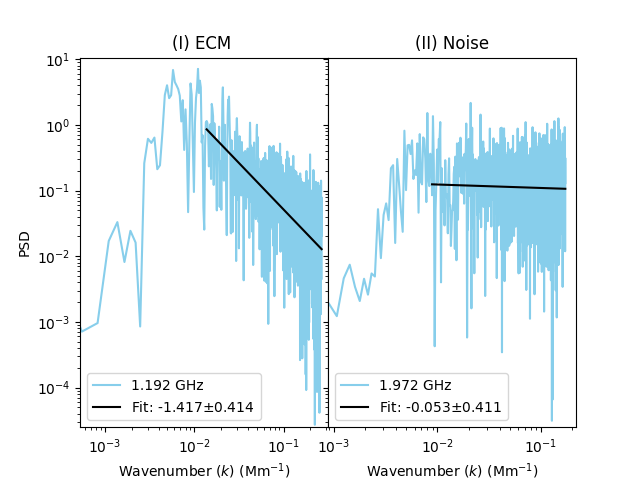}} \\
(A) PSD of $T_B$ time series \\
    \resizebox{58mm}{!}{
\includegraphics[trim={0.0cm 0cm 0.0cm 0.0cm},clip,scale=0.4]{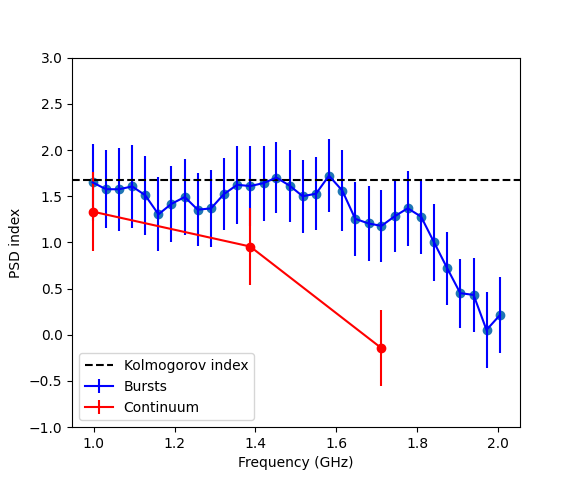}} \\
 (B) Frequency dependence of the PSD index
    \end{tabular}
    \caption{Panel A: The PSD distribution and fit of the fine structures of the time series. The curves in panels (I) and (II) correspond to the wavenumber distribution of $\approx$1.2 GHz and $\approx$2.0 GHz, respectively. The solid black lines are the linear fit to the tail part of the distribution. Note that the index for the 1.2 GHz tail is $\approx$-3/2. Panel B: The linear fit index to the PSD tail is plotted with a function for frequency bands. }
    \label{fig:psd}
\end{figure}

\subsection{Sub-Seconds Fine Structures}
Power Spectral Density (PSD) describes energy transfer from the large to small eddies in a turbulent medium, for example, in solar winds \citep{Jatenco1994}.
To quantify the sub-second features, we pass running median subtracted time-series to PSD, which captures the variability scales in the signal. Since we want to characterize sub-second features, we choose a window size of 3 sec for the running median subtraction. The length scales for the wavenumber ($k$) are obtained by inversely scaling sampling timescales by Alfven velocity corresponding to the gyro frequency layer for a given frequency, i.e. wavenumber $k = t_s/v_A$, where $t_s$ is the timescale. We vary $t_s$ from 50 ms to 3 sec and $v_A\approx$ 16-18 Mm/s corresponding to 1.5 GHz and 1 GHz gyro-frequency layer, respectively, for all 32 frequency channels.
Figure \ref{fig:psd} (A) shows the PSD of the running median subtracted time series of the entire time series, including burst and continuum emission times for two frequencies $\approx$1.2 GHz and $\approx$2.0 GHz. We fit a power-law to the PSD between $10^{-2}$ Mm$^{-1}<k<2\times10^{-1}$ Mm$^{-1}$. We note that for the time-series below 1.5 GHz, the power law fit of the PSD is close to 1.42$\pm$0.41. The 2 GHz PSD shows a flat PSD similar to a flat white noise PSD spectrum. The right panel shows the power-law fits of the PSD for all the frequencies. For the frequencies where radio bursts are seen, the power-law remains flat to near 1.6. Within the error bars, all of them follow the Kolmogorov index (5/3). The 1 GHz continuum emission also shows a similar power law as the radio bursts, while 1.5 GHz and 1.7 GHz deviate away and flatten, resembling the white noise.

\subsection{Wait-time distribution}
\label{subsec:wait}

The FT structures can arise from a self-organised system governed by self-organised criticality (SOC). Wait-time distribution of the events is considered to be a measure of SOC systems. \cite{Aschwanden2021} calculates the power-law index for the wait-time distribution for the SOC processes to be $\approx 2$. We compute the wait-time distribution for the running median subtracted time series with a smoothing window of 20 sec (e.g. a frequency slice in Fig. \ref{fig:sos} (A)). To compute the wait time, we need to define an event. We assume a qualitative definition of features above 1 $\sigma_{run}$ to constitute an event. Here, $\sigma_{run}$ is the standard deviation of the full median subtracted time series.
We compute the wait-time distribution for 1 GHz, 1.5 GHz and 2 GHz for a qualitative estimation. Waiting times are calculated from the start time of the event. Fig. \ref{fig:sos} (C) plots the wait-time distribution for the three frequencies. We note that the distributions of wait times for just noise (2.0 GHz) are slightly steeper than 2, while 1 GHz and 1.5 GHz plots are consistent with a slope of 2. Therefore, these exponential wait times imply that the timescale of the emission cascade is smaller than the time interval of the next event, or in general terms, ``no memory of prior events". Assuming the event is caused by the next bunch of incoming electrons in the masing volume, i.e., no delays and instant emission in the event occurrence. Such occurrences are more probable in multiple masing cells rather than a single cell. One possibility is that many cells can carry multiple electron bunches, and we would observe a combined emission from these cells within each 50 ms time. However, a more detailed and focused study is required to establish this possibility.

\section{Discussion}
\label{sec:discussion}
Imaging and spectroscopy of the VLA radio bursts offer new insights into the ECM emission, particle acceleration, trapping and turbulence in the magnetic loops. Our analysis supports the following evidence of the existence of multiple phenomena like electron injection, electron trapping in the magnetic loops, conducive conditions for ECM instability, periodically coupled electron diffusion into loss-cone and growth of maser source, and modulation of emission by turbulence. We also constrain the energies and pitch angles of the electrons that produce the ECM, which is consistent with the standard flare model. The major findings from the study are as follows based on the analysis in section \ref{sec:model_ecm}, \ref{sec:pulsations} and \ref{sec:fine}, respectively.

\begin{enumerate}
    \item Energetic electrons ($\approx$4-8 keV) from eruptions on the eastern leg are trapped in the over-arching loops and cause a time delay of ECM emission over the sunspot from EUV emission.
    \item The electron acceleration from the solar flare perturbs the continuously operating ECM, resulting in QPPs caused by oscillatory wave-particle energy exchange in the loss cone.
    \item The turbulence contributes to the sub-second temporal structures in the ECM, and the arrival times of the ECM emission are consistent with a self-organised system. 
\end{enumerate}

These results from spectroscopic imaging multi-wavelength data have provided a deeper understanding of the origins of long-lasting radio sources and pulsations. We discuss the implications of the findings in the following sub-sections.

\begin{figure*}
    \centering
        \resizebox{140mm}{!}{
\includegraphics[trim={0.0cm 0cm 0.0cm 0.0cm},clip,scale=0.6]{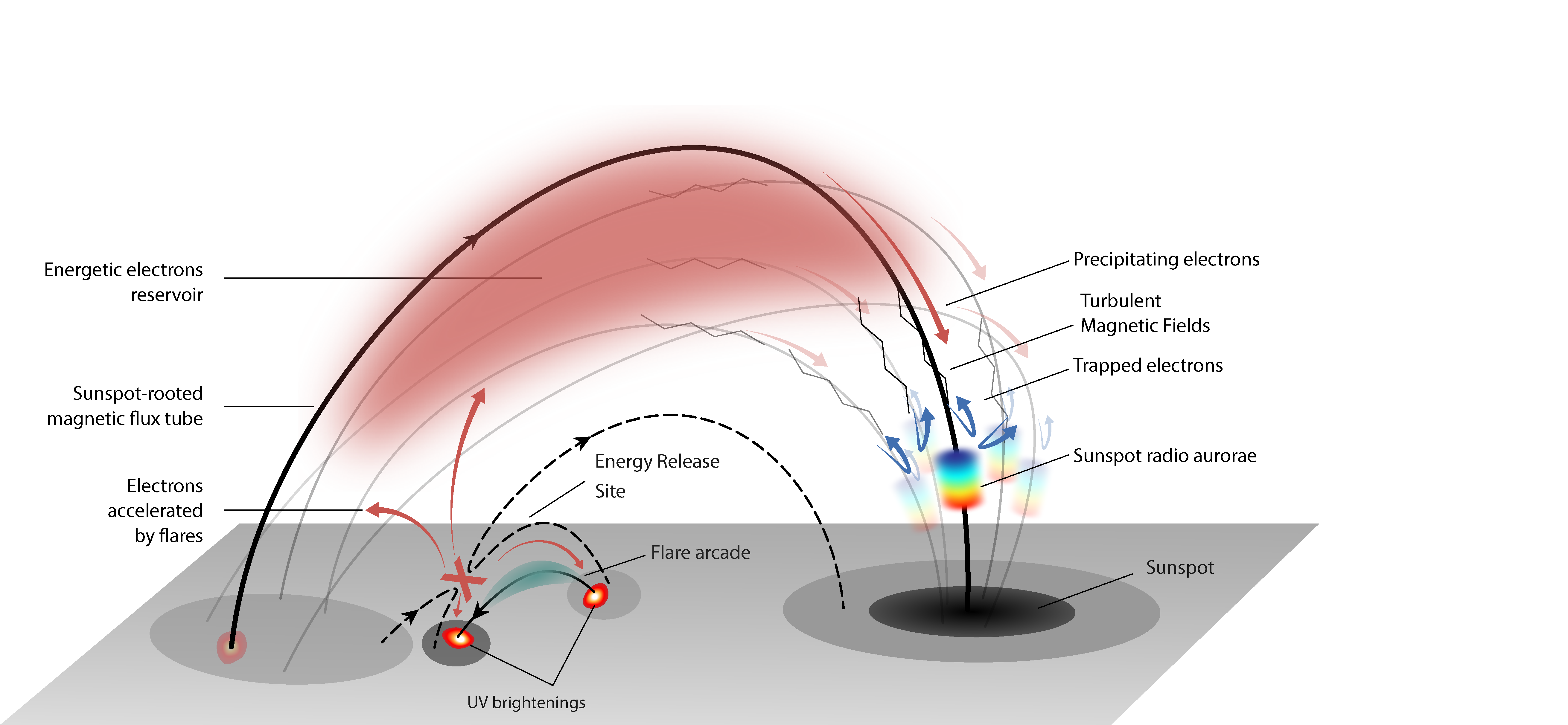}}
\caption{Cartoon showing the ECM masing volume (sunspot radio aurorae) in the solar loop and connection with the solar flare. The figure is adopted and modified from \cite{Yu2023} to include turbulent magnetic field lines shown by zig-zag lines.}
    \label{fig:cartoon}
\end{figure*}



\subsection{Physical Picture of features in ECM}
\label{subsec:physical}
\cite{Yu2023} established the model for the over-arching reservoir for the auroral-like ECM emission over the sunspot. Fig. \ref{fig:cartoon} shows the schematic of the ECM phenomenon, which includes modification from the physical picture described in \citep{Yu2023} based on the current study. Our electron dynamics model and extrapolation show the possibility of an electron reservoir formation and long-lasting converging topology that can inhibit the ECM source for many hours \citep{Yu2023}, and ECM operate continuously. Note that electron reservoir means trapped electrons with a broad range of energies and pitch angles. Given a consistent supply of energetic electrons, the trapped electrons near the loop top can form a reservoir of electrons. The mostly positively charged magnetic region AR12529 remains persistently active on 16 April 2014 (Fig. \ref{fig:aia} (B)). It fluctuated around the B-class flare level and produced 4 C-class bursts in 7 hours before the 18:46 UT burst, suggesting a constant reconnection activity in this AR providing regular intervals of energetic particle production. Thus,  sustained availability of energetic electrons from the eastern loop leg in the connecting loop is ensured. A subset of these energetic electrons travel downwards to produce EUV bright footpoints are seen in Fig. \ref{fig:aia} per the standard flare model. Under suitable conditions, the accelerated electrons travelling upwards can get trapped near the bigger loop top. Untrapped particles precipitate into the sunspot. 

\begin{figure}
    \centering
        \resizebox{85mm}{!}{
\includegraphics[trim={0.0cm 0cm 0.0cm 0.0cm},clip,scale=0.6]{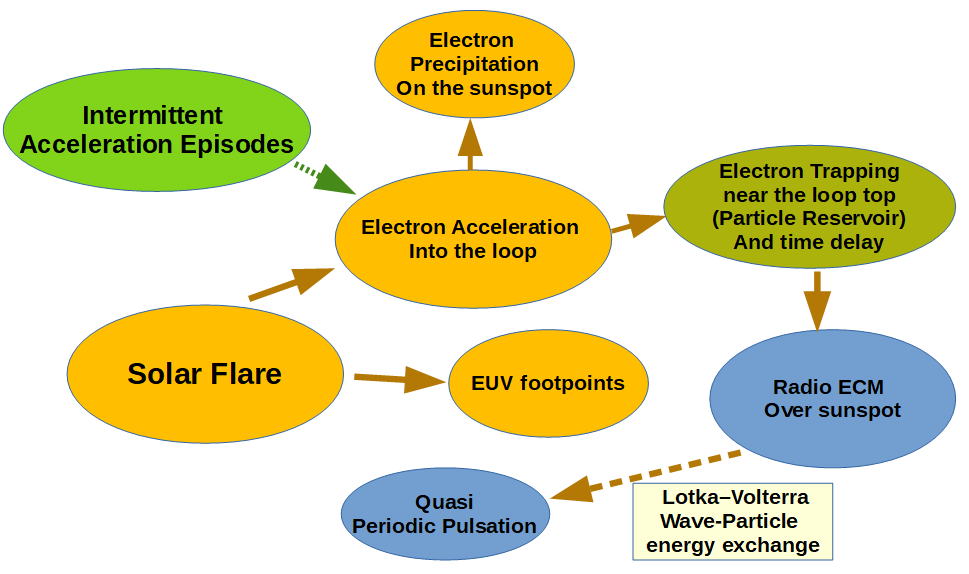}}
    \caption{Flowchart of the particle (electron) dynamics involved in producing the ECM emission. The orange boxes show the electron acceleration and transport sequence. The light green box shows the intermediate low energetic events. The electrons get trapped (green box) at the loop top, and produce long-lasting ECM and QPPs (blue). QPPs are the result of ECM saturation in the wave-particle exchange.} 
    \label{fig:flowchart}
\end{figure}

The reservoir of energetic electrons can slowly and constantly diffuse into the loss cone driven by coulomb collisions and turbulence. The trapping time is determined by magnetic mirroring and collisions, causing a time delay (the dark green block in Fig. \ref{fig:flowchart}). The electrons enter the loss cone and undergo ECM instability at the sunspot leg of the loop to produce ECM. The temporal profile resemblance between the low-energy X-ray and radio bursts with time delay hinted towards electron transport effects and trapping, which are quite well studied with X-ray observations and modelling \citep[e.g.][etc.]{Aschwanden1997,Li2022}. 
In solar flares, the low-energy X-ray peak fluxes are also known to correlate with the high-energy peak fluxes \citep{Isola2007}. Therefore, the temporal association of thermal X-ray fluxes and nonthermal ECM is possible. Even though we compute the $E_{low}\approx$13 keV cut-off FERMI X-ray fits, the nonthermal population with $<13$ keV could still be present, but it could be obscured due to strong thermal background and instrumental limitations.
Based on the location, timing delays and magnetic topology of the observations, a plausible sequence of the electron dynamics for the ECM is shown in Fig. \ref{fig:flowchart}. 

An occasional surge of new electrons from the reservoir is released from a solar flare. These electrons perturb the ECM, producing QPPs. This effect increases ECM growth rate due to perturbed electron velocity distribution \citep{Aschwanden1988I}, the ECM maser can operate in an unsaturated regime. 
Therefore, the overall electron dynamics consists of electron injection into the connecting loop, electron mirroring, which forms the trap and electron reservoir, coulomb collision, and turbulence, which knocks the electrons into the loss cone. 

Another way of producing QPPs can be due to a delay in triggered multiple reconnections leading to the secondary acceleration sites. For e.g. \cite{Battaglia2021} showed radio emissions originating from secondary acceleration sites. Such processes can produce non-co-spatial multiple brightenings but are associated in time. Here, the Fermi/GBM X-ray curve does show a smaller peak before the flare peak.  However, we do not observe any clear signature of secondary acceleration in EUV lightcurves and images.


-

\subsection{Turbulent and Stochastic nature of the ECM}
\label{subsec:dis4}
We note evidence of turbulence and ``self-criticality" in producing fine structures and a variety of FT features (Fig. \ref{fig:psd} ).  Wave-particle interactions are non-linear processes, and exponential growth of mode can cause rapid amplitude fluctuation. Since the growth rate is contributed by the density gradient in the loss cone, a minor fluctuation of the electron distribution affects the growth rate linearly, but the wave energy exponentially \citep[e.g.][]{Melrose1982,Aschwanden1988I}. Therefore, in a turbulent inhomogeneous medium, the wave energies would intrinsically have a complex spatial and temporal pattern, which can result in the observed FT features.
The ECM growth rate is driven by persistent nonthermal electron populations, i.e. excitation and quenched by quasi-linear relaxation in the velocity space, i.e. relaxation, respectively. Their interplay can produce regular spatial-spectral-temporal patterns, unlike the random pattern. The entire system can become ``self-similar", where the emission is amplified not by external nonthermal electrons but by fluctuations amplified by the feedback like a non-linear relaxation. Such a system can also become ``critical" (SOC), for example, if continuously driven by nonthermal electrons and relaxed randomly, not episodically \citep{Aschwanden2021}. In such case, the radio emission would continuously show FT features and slopes of the wait time distribution $\approx$2 (Fig.\ref{fig:sos} (C)), i.e. a fast emission cascade (Sec. \ref{subsec:wait}). One plausible scenario could be the presence of multiple masing cells forming a combined coherent FT structure in the ECM source. A similar power-law slope of $\alpha \approx$ 2.1–2.4 was found in the soft X-rays fitted with a non-stationary Poisson process  \citep[e.g.][]{Wheatland2000}.
Similar behaviour can be reproduced with a shell model of turbulence \citep{Boffetta1999}, or with a Levy function \citep{Lepreti2001}.
Therefore, the wait-time power law's uniqueness is ambiguous and makes it hard to classify it as SOC-dominated, turbulent, or both. In the present case, the consistency in $L_{drift}$ and loop top extrapolation width, the Kolmogorov PSD distribution of the ECM fine structures over turbulent scales ($l_{turb}$) between 5 Mm $\leq l_{turb} \leq$100 Mm (Fig. \ref{fig:psd}), and the presence of FT structures support both the turbulence and SOC nature in the ECM.

\section{Conclusion}
\label{subsec:conclusion}
Using a comprehensive multi-wavelength investigation and modelling in the spatio-temporal-spectral domain, we studied the nature of long-lasting auroral-like ECM emission over the sunspot, over-arching electron reservoir for ECM and the role of the accelerated nonthermal electrons originating in a solar flare in ECM emission. We constrain the emission mechanism to be ECM at the second harmonic (s=2) o-mode. At EUV and X-ray wavelengths, a solar flare was seen at the eastern edge of the active region, while the ECM source was persistently present at the western edge over the sunspot. Several hot magnetic loops connecting the flare site to the sunspots are seen in AIA 171 \AA, and modelled by magnetic extrapolation. By simple modelling of electron dynamics in the magnetic loop, we proposed trapping of accelerated electrons injected from the active eastern loop leg, forming a loop top electron reserve, which is consistent with the standard flare model. Using X-ray spectroscopic analysis, we obtained energies and densities of thermal and nonthermal electron populations during the flare. Radio bursts with 5-second period QPPs were seen with a 38-second time delay w.r.t. the solar flare. The X-ray fits capture the higher end of the accelerated population, $E_{low}>13keV$, while the lower end is responsible for the ECM emission. Using the X-ray analysis and delay in ECM emission, we estimate the energies between $\approx$4-8 keV form the electron reservoir with varied pitch angles $\alpha_0>26^o$. As \cite{Yu2023} showed that this ECM source lasts for many hours, we find that at smaller second and sub-second temporal scales, the emission is governed by loss cone diffusion supported by collisions and magnetic turbulence with Kolmogorov power-law distribution over $\approx$5 Mm  $\leq l_{turb} \leq$100 Mm length scales. Fulled by over-arching electron reservoir and loss-cone diffusion, the ECM operates continuously and sometimes shows coherent and self-similar FT structures. The additional electrons from intermittent solar flare saturate the ECM emission and produce 5-second period QPPs, otherwise absent in the continuous phase of ECM.


Overall, long-lasting ECM sources are viable under conducive magnetic geometry and electron acceleration sites and can help us develop more detailed particle transport models. The fine structures offer a more detailed view of the emission mechanism and coronal medium. Therefore, more observations and analysis of long-lasting radio sources are required using a multi-wavelength approach. In addition, combined radio observations with solar-based observatories like Parker Solar Probe and Solar Orbiter with AIA can help get stereoscopic views. This would allow us to constrain the particle propagation and emission more precisely.


\begin{acknowledgments}

This work was supported by the Swiss National Science Foundation (grant no. 200021\_175832). We acknowledge and thank the referee for the critical review of the manuscript. It makes use of public archival VLA data from the observing program VLA/16A-377	 carried out by the National Radio Astronomy Observatory (NRAO). The authors acknowledge Tim Bastian, Dale Gary, and Stephen White for their help in carrying out the observing program and Gregory Fleishman for help with the GX-simulator package. S.Y. is supported by NASA grants
80NSSC20K1283/SV0-09025 and 80NSSC21K0623 to NJIT. 
Y.L. and B.C. are supported by NSF grant AGS-1654382 to the New Jersey Institute of Technology.
The NRAO is a facility of the National Science Foundation (NSF) operated under cooperative agreement by Associated Universities, Inc.
\end{acknowledgments}

\appendix

\bibliography{manuscript}{}

\end{document}